\documentclass[aps]{revtex4}
\usepackage{amsfonts}
\usepackage{amsmath}
\usepackage{amssymb,epsf}
\usepackage{color}

\begin{document}

\title{Thermodynamics and and phase transition of topological dilatonic
Lifshitz-like black holes}
\author{S. H. Hendi $^{1,2,3}$ \footnote{
email address: hendi@shirazu.ac.ir}, F. Azari$^{1,2}$, E.
Rahimi$^{1,2,4}$, M. Elahi$^{1,2}$, Z. Owjifard$^{1,2}$ and Z.
Armanfard$^{5}$} \affiliation{$^{1}$Department of Physics, College
of Sciences, Shiraz University, Shiraz 71454, Iran\\
$^{2}$Biruni Observatory, College of Sciences, Shiraz University,
Shiraz 71454, Iran\\
$^{3}$Canadian Quantum Research Center 204-3002 32 Ave Vernon, BC
V1T 2L7 Canada\\
$^{4}$Department of Elementary Particles, Faculty of Physics,
University of Kashan, Kashan, Iran\\
$^{5}$ Department of physics and astronomy, Washington State
University, Pullman, Washington 99164-2814, USA}

\begin{abstract}
It is known that scalar-tensor gravity models can be studied in Einstein and
Jordan frames. In this paper, we consider a model of scalar-tensor gravity
in Einstein's frame to calculate the Lifshitz-like black hole solutions with
different horizon topologies. We study thermodynamic properties and first
order van der Waals like phase transition, and find that the Lifshitz
parameter affects the phase structure. In addition, we investigate thermal
stability by using the behavior of heat capacity and various methods of
geometrical thermodynamics.
\end{abstract}

\maketitle

\section{Introduction}

Regarding the successful consequences of the general relativity,
especially the gravitational waves \cite{LIGO} prediction, it may
be believed that it is a fundamental theory of gravitation.
Nonetheless, it has recently been found that GR is neither a
renormalizable theory in small-scale (UV regime) nor can explain
the late time acceleration of the universe in the large-scale (IR
regime). Nowadays, it is a conviction that GR needs a UV-complete
gravitational theory and an IR one as well.

Taking into account a Lagrangian with a supplementary higher order
derivatives of the metric, one may solve the problem of
renormalizability in UV scale. Although some higher derivative
modifications of GR provide the renormalizability, their
applications may lead to ghosts. This is due to the fact that
although the higher spatial derivatives can improve
renormalizability, the higher time derivatives will lead to
ghosts. Accordingly, one should look for a Lorentz violation
theory in such a way that the higher spatial derivatives will be
decomposed from higher time ones. In this regard, the
Horava-Lifshitz \cite{HL1,HL2} approach helps us to build a theory
with an anisotropic scaling between time and space
\begin{equation}
t\rightarrow \lambda ^{z}t,\;\;\;\;\;\;\;\;\;\;\;\;\;\;\;\mathbf{x}%
\rightarrow \lambda \mathbf{x,}  \label{HL-trans}
\end{equation}%
where $z$ is the dynamical critical exponent. Considering Eq. (\ref{HL-trans}%
), one finds that although the Galilean invariance is preserved,
isotropy and Lorentz symmetry between space and time will be
broken. It is also notable that $z\geq n$ ($n$ denotes the spatial
dimension of the spacetime) is a requirement for the
power-counting renormalizability, while the Lorentz invariance can
be recovered for $z=1$ \cite{HL1,HL2,PCR2,PCR3,PCR4,PCR5,PCR6}.

With this in mind, we examine the effect of the critical exponent
($z$) on the thermodynamics of black holes, a theme that has
received considerable attention in recent decades. More
specifically, we focus on the possible phase transition of
topological Lifshitz-like black holes. Among different classic and
quantum aspects of black holes, investigation of thermodynamic
properties has always played a central role in the black hole
physics. As a window to quantum gravity, the role of black hole
thermodynamics, phase transition and criticality cannot be denied.
Thermodynamic description of black holes refers to the primitive
works of Bekenstein and Hawking on the entropy and temperature of
the event horizon \cite{Beck,Haw1,Haw2}. Strictly speaking, they
showed that the area of event horizon and its surface gravity are,
respectively, associated with the entropy and temperature of the
black hole. Regarding the total energy (mass) of the black hole as
a conserved charge associated with the time-translation symmetry,
we can examine the first law of thermodynamics, just as the
analogy with the common first law of thermodynamics in usual
systems. In a realistic scenario, we should check the robustness
of black hole with respect to small perturbation. In other words,
we have to study both dynamic and thermodynamic stabilities of the
solutions. Dynamical stability can be investigated through the
quasi-normal modes of the solutions, while thermodynamic stability
may be studied based on the response of a system to the
fluctuations of energy, temperature and other thermodynamic
parameters near the equilibrium. Based on perturbative variables,
thermodynamic stability of a black hole can be studied via
micro-canonical, canonical and grand canonical ensembles. In
recent years, black hole thermodynamics and the criteria of
thermal
stability have been investigated in literature \cite%
{BHthermo1,BHthermo2,BHthermo3,BHthermo4,BHthermo5,BHthermo6}.
Regarding a typical black hole as a thermodynamic system, one may
look for thermodynamic variables as the first step. In addition to
the first law of thermodynamics, the Smarr relation with scaling
approach can be helpful in this regard. Based on the Smarr
relation of black holes, one has to modify the first law of
thermodynamics in the so-called extended phase space. Working in
the extended phase space, the cosmological constant interpreted as
a dynamical pressure, and AdS charged black holes undergo a van
der Waals like phase transition.

Geometrical thermodynamics (GT) is one of the interesting methods to
investigate the black hole phase transition. This approach was employed by
the works of Gibbs \cite{Gibbs} and Caratheodory \cite{Caratheodory}. Based
on this method one could hire thermodynamical potentials and their
corresponding quantities to define a phase space. Constructing the
thermodynamical metric, the divergence points of its Ricci scalar provide
important information related to the possible phase transition of black hole
system. The most common thermodynamical metrics are Weinhold, Ruppeiner,
Quevedo and HPEM. The first one which is introduced by Weinhold \cite%
{Weinhold,Weinhold01} is defined on the space of equilibrium state of
thermodynamic systems. After that Ruppeiner defined another metric in $1979$
\cite{Ruppeiner,Ruppeiner01}. The shortcomings of both Weinhold and
Ruppeiner metrics come from the fact that they are not invariant under
Legendre transformation. Since Legendre invariance plays a major role in GT
method, Quevedo proposed the first Legendre invariance metric \cite%
{Quevedo1,Quevedo11} which solved some problems of previous
methods. However, Quevedo metric is not, completely, a successful
model in several specific systems and its Ricci scalar has extra
divergence point without physical interpretation. Finally, a new
metric was proposed \cite{HPEM,Hendi01,Hendi02} in which the
problem of mismatched divergency is not observed
\cite{Soroushfar,Vetsov,Chabab}.

This paper is organized as follow. At first, the field equations and the
static solutions of the Lifshitz model are presented. Then, the
thermodynamic and conserved quantities are calculated, and thermal stability
and the existence of phase transition are examined by studying the behaviour
of the system through $C_{Q}-r_{+}$, $P-V$, $G-T$ diagrams. Finally, the
geometrical thermodynamic method is used to confirm the results. We end the
paper with some concluding remarks.

%%%%%%%%%%%%%%%%%%%%%%%%%%%%%%%%%%%%%%%%%%%%%%%%%%%%%%%%%%%%%%%%%%%%%%%%%%%%%%%%%%%%%%%%%%%%%%%%%%%%%%%%%%%%%

\section{Exact solutions with their geometric and thermodynamic properties
\label{FE}}

Here, we consider a model of scalar-tensor gravity in which the scalar field
is non-minimally coupled with the Maxwell field as \cite%
{Alishah,Main,Main2,Main3}
\begin{equation}
I=\int_{\mathcal{M}}d^{4}x\sqrt{-g}\left[ R-2\Lambda -\frac{1}{2}\partial
_{\mu }\phi \partial ^{\mu }\phi +V(\phi )-e^{\lambda \phi }F^{\mu \nu
}F_{\mu \nu }\right] ,  \label{Action}
\end{equation}%
where
\begin{equation}
V(\phi )=V_{0}e^{\gamma \phi },
\end{equation}%
and $\Lambda $, $\lambda $, $\gamma $ and $V_{0}$ are some constant
parameters of the theory. The field equations of the above action can be
obtained via the variational principle, as%
\begin{eqnarray}
R_{\mu \nu }+\frac{1}{2}\left[ V(\phi )-2\Lambda \right] g_{\mu \nu } &=&%
\frac{1}{2}\partial _{\mu }\phi \partial _{\nu }\phi +2e^{\lambda \phi
}\left( F_{\mu \sigma }F_{\nu }{}^{\sigma }-\frac{g_{\mu \nu }}{4}F_{\alpha
\beta }F^{\alpha \beta }\right) ,  \label{GravityEq} \\
\nabla ^{2}\phi &=&-\frac{dV(\phi )}{d\phi }+\lambda e^{\lambda \phi
}F_{\alpha \beta }F^{\alpha \beta },  \label{PhiEq} \\
\nabla _{\mu }\left( e^{\lambda \phi }F^{\mu \nu }\right) &=&0.
\label{MaxEq}
\end{eqnarray}%
Now, we consider the following ansatz for the metric,%
\begin{equation}
ds^{2}=-\left( \frac{r}{r_{0}}\right) ^{z}B(r)dt^{2}+\frac{1}{B(r)}%
dr^{2}+r^{2}d\Omega _{k}^{2},  \label{metric}
\end{equation}%
where the line element $d\Omega _{k}^{2}$ is the metric of two-dimensional
(unit) hypersurface with constant curvature $6k$ and volume $V_{k}$ with the
following explicit form
\begin{equation}
d\Omega _{k}^{2}=\left\{
\begin{array}{cc}
dx_{1}^{2}+{{\sin }^{2}}{x_{1}}dx_{2}^{2},\; & k=1 \\
&  \\
dx_{1}^{2}+{\sinh ^{2}}{x_{1}}dx_{2}^{2},\; & k=-1 \\
&  \\
dx_{1}^{2}+dx_{2}^{2},\; & k=0%
\end{array}%
\right. ,
\end{equation}%
in which different values of the topological factor ($k=-1,0,+1$) indicate
the surface (event horizon) topology as positive (spherical, ${\mathbb{S}^{2}%
}$), zero (planar, $\mathbb{R}^{2}$), or negative (hyperbolic, $\mathbb{H}%
^{2}$). \bigskip In order to solve the field equations, we assume that $\phi
=\phi (r)$ and the only nonzero components of the gauge field are $%
F_{rt}=-F_{tr}\neq 0$. As the first step, we consider the Maxwell equation (%
\ref{MaxEq}) with the subtraction of the $tt$ and $rr$ components of the
Einstein equation ($R_{tt}-R_{rr}$) to obtain%
\begin{eqnarray}
e^{\phi } &=&\left( {\frac{r}{r_{0}}}\right) ^{\sqrt{2z}},  \label{Phi} \\
F_{tr} &=&\frac{q}{r^{2}}\left( \frac{r_{0}}{r}\right) ^{\lambda \sqrt{2z}-%
\frac{z}{2}}.  \label{Ftr}
\end{eqnarray}

Now, we can obtain the metric function $B(r)$ by utilizing the other
components of the Einstein equations (\ref{GravityEq}), as%
\begin{equation}
B(r)=\frac{2k}{z+2}-\frac{m}{r^{1+\frac{z}{2}}}+\frac{V_{0}r^{2}}{\Gamma }%
\left( \frac{r}{r_{0}}\right) ^{\gamma \sqrt{2z}}+\frac{2q^{2}}{\Xi r^{2}}%
\left( \frac{r_{0}}{r}\right) ^{\lambda \sqrt{2z}}-\frac{2\Lambda r^{2}}{6+z}%
.  \label{Sol}
\end{equation}

In these relations, $z>0$ is the dimensionless Lifshitz parameter, $r_{0}$
is an arbitrary constant with length dimension, $\Gamma =2\gamma \sqrt{2z}%
+6+z$, $\Xi =2\lambda \sqrt{2z}+2-z$, $m$ is an integration constant related
to the mass and $q$ is the electric charge of the electromagnetic field
(Hint: comparing the mass term with the mass term of $4-$dimensional
Schwarzschild-like black holes ($\frac{-m_{0}}{r}$), one finds $%
m=m_{0}r_{0}^{z/2}$ as a relation between the integration constant, $m$, and
the geometrical mass, $m_{0}$).

\begin{figure}[tbp]
$%
\begin{array}{ccc}
\epsfxsize=5.4cm \epsffile{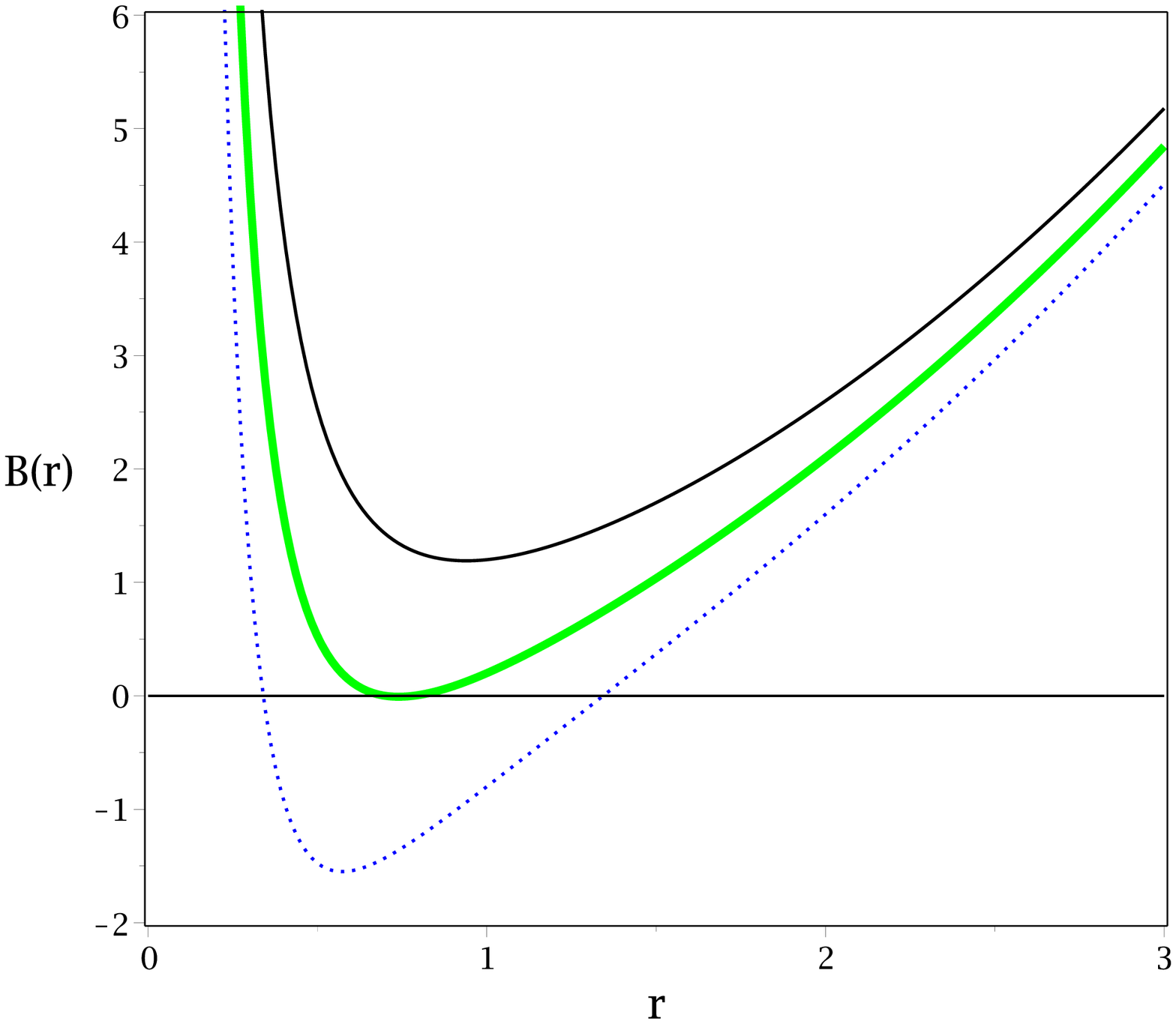} & \epsfxsize=5.4cm %
\epsffile{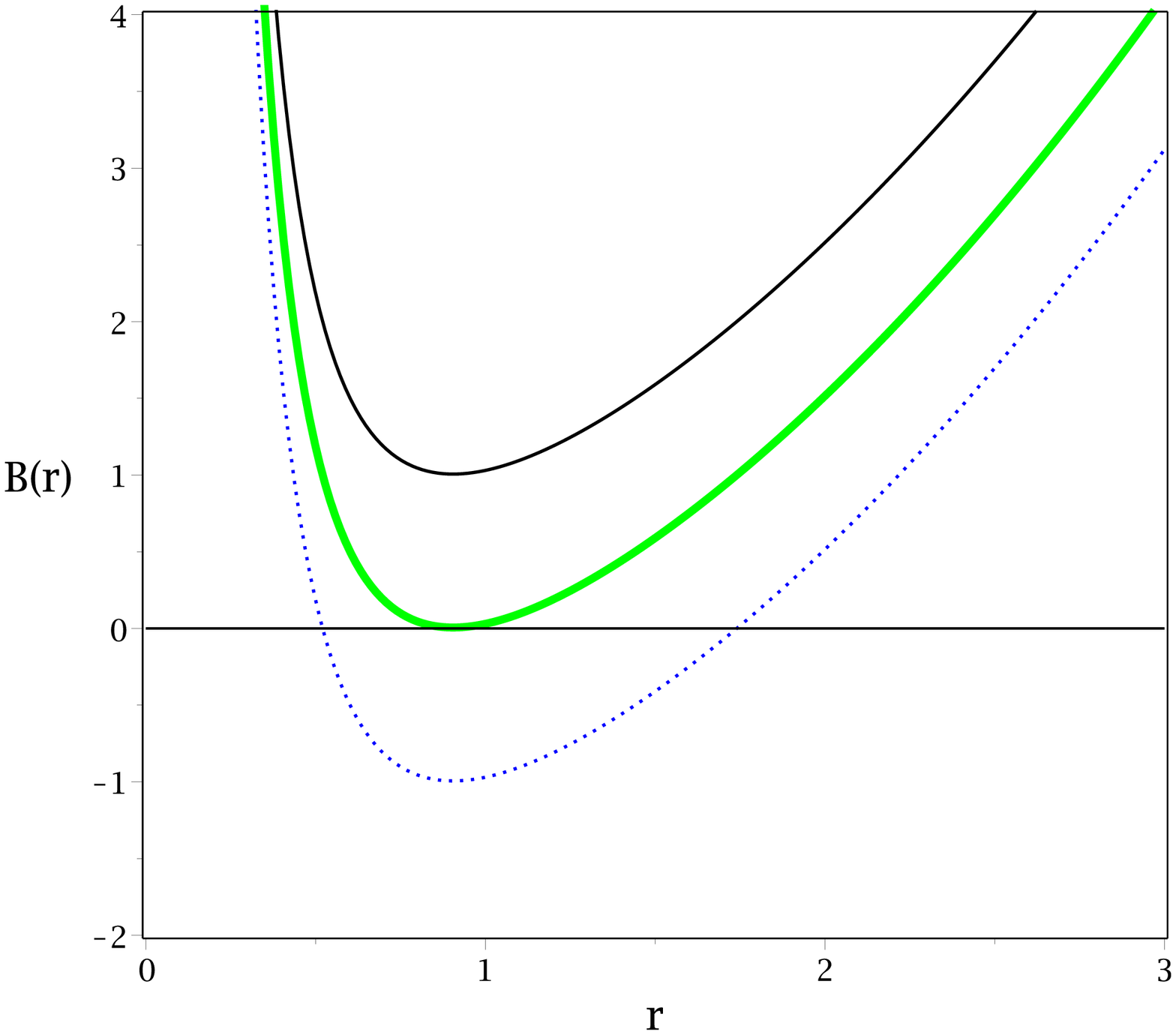} & \epsfxsize=5.8cm \epsffile{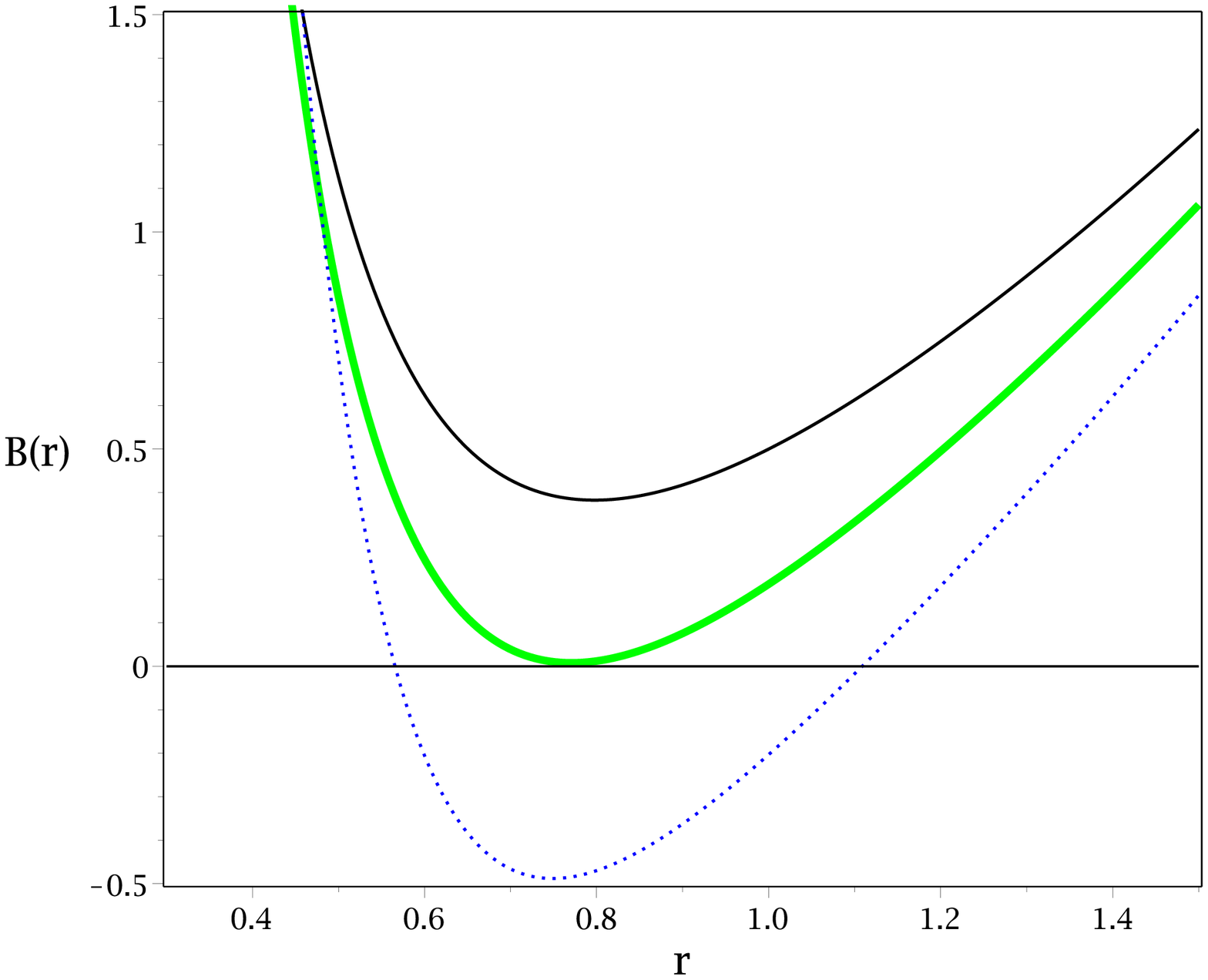}%
\end{array}
$%
\caption{$B(r)$ versus $r$ for $\protect\lambda=1$, $\protect\gamma=1$, $q=1$%
, $\Lambda=-1$, $V_{0}=1$, $r_{0}=1$.\newline
\textbf{Left panel:} $z=0$, $k=1$, $m=1.3$ (solid line), $m=2.3$ (bold line)
and $m=3.3$ (dotted line).\newline
\textbf{Middle panel:} $z=0$, $m=1.47$, $k=1$ (solid line), $k=0$ (bold
line) and $k=-1$ (dotted line).\newline
\textbf{Right panel:} $m=2$, $k=1$, $z=0$ (solid line), $z=0.07$ (bold line)
and $z=0.5$ (dotted line). }
\label{MetricFig}
\end{figure}

In order to find the behavior of the curvature, we should
calculate the curvature scalars. Calculating the Ricci scalar and
Riemann invariant (Kretschmann scalar), we obtain
\begin{eqnarray}
R &=&-B^{\prime \prime }(r)-\frac{(3z+8)}{2r}B^{\prime }(r)-\frac{%
(z^{2}+2z+4)}{2r^{2}}B(r)+\frac{2k}{r^{2}},  \label{R} \\
&&  \notag \\
R_{\alpha \beta \gamma \delta }R^{\alpha \beta \gamma \delta } &=&B^{\prime
\prime 2}(r)+\frac{z(z-2)}{r^{2}}\left( B(r)+\frac{3rB^{\prime }(r)}{z-2}%
\right) B^{\prime \prime }(r)+  \notag \\
&&\frac{(9z^{2}+16)}{4r^{2}}B^{\prime 2}(r)+\frac{z(3z^{2}-6z+8)B(r)}{2r^{3}}%
B^{\prime }(r)+  \notag \\
&&\frac{(z^{4}-4z^{3}+12z^{2}+16)}{4r^{4}}B^{2}(r)-\frac{8k}{r^{4}}B(r)+%
\frac{4k^{2}}{r^{4}}.  \label{RR}
\end{eqnarray}%
Inserting the functional form of $B(r)$ into Eqs. (\ref{R}) and
(\ref{RR}), we obtain
\begin{eqnarray}
R &=&b_{1}kr^{-2}+b_{2}mr^{-3-\frac{z}{2}}+b_{3}q^{2}r^{4-\lambda \sqrt{2z}%
}+b_{4}V_{0}r^{\gamma \sqrt{2z}}+\frac{\left( z^{2}+8z+24\right) }{6+z}%
\Lambda \\
&&  \notag \\
R_{\alpha \beta \gamma \delta }R^{\alpha \beta \gamma \delta } &=&\alpha
_{1}k^{2}r^{-4}+\alpha _{2}kmr^{-\frac{z}{2}-5}+\alpha
_{3}m^{2}r^{-z-6}+\alpha _{4}V_{0}^{2}r^{2\gamma \sqrt{2z}}+  \notag \\
&&\alpha _{5}kV_{0}r^{\gamma \sqrt{2z}-2}+\alpha _{6}mV_{0}r^{2\gamma \sqrt{%
2z}-\frac{z}{2}-3}+\alpha _{7}q^{4}r^{-2\lambda \sqrt{2z}-8}+  \notag \\
&&\alpha _{8}kq^{2}r^{-\lambda \sqrt{2z}-6}+\alpha _{9}mq^{2}r^{-\frac{z}{2}%
-\lambda \sqrt{2z}-7}+\alpha _{10}V_{0}q^{2}r^{(\gamma -\lambda )\sqrt{2z}%
-4}+  \notag \\
&&\alpha _{11}\Lambda kr^{-2}+\alpha _{12}\Lambda mr^{-\frac{z}{2}-3}+\alpha
_{13}\Lambda V_{0}r^{\gamma \sqrt{2z}}+\alpha _{14}\Lambda q^{2}r^{-\lambda
\sqrt{2z}-4}  \notag \\
&&+\frac{\Lambda ^{2}}{(6+z)^{2}}\left[ (z^{2}+4z+16)(z+4)z+96\right] ,
\end{eqnarray}%
where $b_{i}$'s and $\alpha _{i}$'s are some constants related to $z$, $%
\gamma $ and $\lambda $. The last (constant) term of this equation indicates
that, in general, the asymptotic behavior of the solutions is AdS with an
effective cosmological constant. However, considering the coefficient "$(%
\frac{r}{r_{0}})^{z}$" in the metric, one finds that the solutions are
asymptotically Lifshitz-like solutions. We also find that the the curvature
invariants are finite everywhere except for $r=0$. In other words, there is
a curvature singularity located at $r=0$, which is simply a sign that the
spacetime may contain a singular black hole, a singularity covered by an
event horizon. Taking into account Eq. (\ref{Sol}) and Fig. \ref{MetricFig},
it is clear that the singularity can be covered with an event horizon, and
therefore, one can interpret it as a black hole. To speak more clearly,
depending on the values of free parameters, the metric function may have two
roots (black hole with an inner Cauchy horizon and an outer event horizon),
an extreme root (extremal black hole) or no root (naked singularity: no
black hole). In addition, according to the Fig. \ref{MetricFig}, we observe
that $\lim_{r\longrightarrow 0^{+}}B(r)=+\infty $, and therefore, the
singularity is timelike and avoidable. It is also notable that for $z=0$,
Eq. (\ref{Sol}) reduces to the Reissner-Nordstr\"{o}m-AdS black hole
solutions.

%%%%%%%%%%%%%%%%%%%%%%%%%%%%%%%%%%%%%%%%%%%%%%%%%%%%%
%%%%%%%%%%%%%%%%%%%%%%%%%%%%%%%%%%%%%%%%%%%%%%%%%%%

Regarding the mentioned geometrical properties of the solutions, we are in a
position to examine the first law of thermodynamics. To do so, we should
calculate the conserved and thermodynamic quantities.

First, we focus on the Hawking temperature. In order to calculate
it, we can use the analytical continuation of the metric or the
surface gravity interpretation. Both methods lead to the following
unique result
\begin{eqnarray}
T &=&\frac{B^{\prime }(r_{+})}{4\pi }\left( \frac{r_{+}}{r_{0}}\right) ^{%
\frac{z}{2}}=\frac{\left( \frac{r_{+}}{r_{0}}\right) ^{\frac{z}{2}}r_{+}}{%
4\pi }\times  \notag \\
&&\left( \frac{k}{r_{+}^{2}}-\Lambda -\frac{q^{2}\left( \frac{r_{0}}{r+}%
\right) ^{\lambda \sqrt{2z}}}{r_{+}^{4}}+\frac{V_{0}\left( \frac{r_{+}}{r_{0}%
}\right) ^{\gamma \sqrt{2z}}}{2}\right) ,  \label{Temp}
\end{eqnarray}%
where we used the fact that $r_{+}$ is the largest real positive root of
metric function, and therefore, $B(r_{+})=0$.

Another related quantity is the entropy. Since we are working with a
Lifshitz-like model, applying the area law is challenging. Respecting the
first law of thermodynamics, we can obtain the entropy by using $\delta
S=\delta M/T$, and therefore, we need to have the finite mass.

Definition of finite mass for asymptotically AdS spacetimes has
been given by Ashtekar-Magnon-Das (AMD) \cite{Mass1,Mass2}. They
used the Penrose conformal completion techniques to map the
boundary at infinity to a finite distance. The conserved
quantities (such as mass and angular momentum in stationary
spacetime) in the AMD approach are given in terms of the electric
part of the Weyl tensor (see \cite{JKMO} for more details). The
electric part of the Weyl tensor is corresponding to the Weyl
tensor projected to the boundary, and one can find the finite
conserved charge for the asymptotically AdS space for a given
boundary conditions. Therefore, it is important to find the
asymptotic (fall-off) behavior of the Weyl tensor in determining
the conserved charges. It is worth mentioning that the Weyl tensor
vanishes identically for the global AdS space, and therefore, the
corresponding finite conserved charge vanishes as well in such a
space. As a result, finite conserved charge in the AMD approach
does not include the vacuum energy of the AdS space. Using the AMD
approach, the total mass (per unit volume of $t=cte$ and $r=cte$
boundary: $V_{k}$) can be written as \cite{Mass1,Mass2,JKMO,Mass3}
\begin{equation}
M=\frac{m_{0}}{8\pi }=\frac{m}{8\pi }r_{0}^{-\frac{z}{2}},
\label{Total-Mass}
\end{equation}%
where $m$ is related to the geometrical mass ($m=m_{0}r_{0}^{z/2}$) of the
metric function which is calculated via $B(r_{+})=0$ as
\begin{equation}
m=\left[ \frac{2k}{z+2}+\frac{V_{0}r_{+}^{2}}{\Gamma }\left( \frac{r_{+}}{%
r_{0}}\right) ^{\gamma \sqrt{2z}}+\frac{2q^{2}}{\Xi r_{+}^{2}}\left( \frac{%
r_{0}}{r_{+}}\right) ^{\lambda \sqrt{2z}}-\frac{2\Lambda r_{+}^{2}}{6+z}%
\right] r_{+}^{(1-\frac{z}{2})}.  \label{Mass}
\end{equation}%
Considering Eq. (\ref{Temp}), (\ref{Total-Mass}) and (\ref{Mass}) with the
validity of the first law ($\delta S=\delta M/T$), one can obtain the
following relation for entropy (per unit volume $V_{k}$)
\begin{equation}
S=\frac{r_{+}^{2}}{4},  \label{Entropy}
\end{equation}%
which is the area law of black hole entropy. Considering the
relation between the Faraday tensor and gauge potential, $F_{\mu
\nu }=\partial _{\mu }A_{\nu }-\partial _{\nu }A_{\mu }$ with Eq.
(\ref{Ftr}), we can find the nonzero component of the gauge
potential is $A_{t}=-\int F_{tr}dr$, and therefore, the scalar
potential $U$ at the event horizon with respect to the reference
($r\longrightarrow \infty $) can be written as
\begin{equation}
U=-\int_{r_{+}}^{\infty }F_{tr}dr=\frac{2q}{\Xi r_{+}}\left( \frac{r_{0}}{%
r_{+}}\right) ^{\lambda \sqrt{2z}-\frac{z}{2}}.  \label{Pot}
\end{equation}

In addition, in order to calculate the electric charge, we have to
consider the projections of the electromagnetic field tensor on
special hypersurfaces. The normal to such hypersurfaces is
\begin{equation*}
u^{0}=\frac{1}{N},\text{ \ \ }u^{r}=0,\text{ \ \ }u^{i}=-\frac{V^{i}}{N},
\end{equation*}%
where $N$ and $V^{i}$ are the lapse function and shift vector. Using the
Gauss law and the fact that the electric field is $E^{\mu }=g^{\mu \rho
}e^{\lambda \phi }F_{\rho \nu }u^{\nu }$, we can obtain the electric charge $%
Q$ (per unit volume $V_{k}$)  by calculating the flux of the
electric field at infinity, yielding
\begin{equation}
Q=\frac{q}{4\pi }.  \label{Charge}
\end{equation}%
\bigskip Now, it is straightforward to confirm the first law of
thermodynamics as
\begin{equation}
dM=TdS+UdQ,  \label{First}
\end{equation}%
since $T=\left( \frac{\partial M}{\partial S}\right) _{Q}$ and $U=\left(
\frac{\partial M}{\partial Q}\right) _{S}$ are, respectively in agreement
with those of calculated in Eqs. (\ref{Temp}) and (\ref{Pot}). As an
important comment, we should note that although the first law of
thermodynamics is satisfied, the corresponding Smarr relation is not valid.
Such an anomaly is one of the motivations of working in the extended phase
space thermodynamics which will be addressed in the next section.

\section{Extended phase space, phase transition and thermal stability}

\subsection{van der Waals like behavior}

Here, we want to work in the extended phase space thermodynamics.
Considering the variation of the cosmological constant (motivated by the
vacuum expectation value of a quantum field), we can study thermal stability
in the extended phase space and look for a possible phase transition. In the
context of extended phase space, we can regard the cosmological constant as
a dynamical pressure $P=-\frac{\Lambda }{8\pi }$ \cite%
{PV1,PV2,PV3,PV4,PV5,PV6}. Here, in order to define a suitable dynamical
pressure, we should compare the $r^{2}-$terms of Eq. (\ref{Sol}), $\frac{%
-2\Lambda r^{2}}{6+z}$, with that of RN-AdS solutions, $\frac{-\Lambda r^{2}%
}{3}$. In other words, we can rewrite the $r^{2}-$terms of Eq. (\ref{Sol})
as $\frac{-2\Lambda r^{2}}{6+z}=\frac{-\Lambda _{\text{eff}}\;r^{2}}{3}$,
and therefore,
\begin{equation}
P=-\frac{\Lambda _{\text{eff}}}{8\pi }=\frac{-3\Lambda }{4\pi (6+z)},
\label{P0}
\end{equation}%
with the thermodynamic volume as its conjugate quantity in the first law of
thermodynamics. Regarding the cosmological constant as a thermodynamic
pressure, we can find the extended version of the first law of
thermodynamics with an additional volume-pressure term in the enthalpy
representation
\begin{equation}
dM=TdS+UdQ+VdP,  \label{ExtendedFirst}
\end{equation}%
where the mass of black hole is interpreted as the enthalpy $(M\equiv H)$,
i.e.,
\begin{eqnarray}
M &=&H=\frac{m}{8\pi }r_{0}^{-\frac{z}{2}}=  \notag \\
&&\frac{r_{+}}{8\pi }\left[ \frac{2k}{z+2}+\frac{V_{0}r_{+}^{2}}{\Gamma }%
\left( \frac{r_{+}}{r_{0}}\right) ^{\gamma \sqrt{2z}}+\frac{2q^{2}}{\Xi
r_{+}^{2}}\left( \frac{r_{0}}{r_{+}}\right) ^{\lambda \sqrt{2z}}+\frac{8\pi P%
}{3}r_{+}^{2}\right] \left( \frac{r_{+}}{r_{0}}\right) ^{\frac{z}{2}},
\label{MassFinal}
\end{eqnarray}%
where we used Eqs. (\ref{Mass}) and (\ref{P0}). Using Eq. (\ref%
{ExtendedFirst}), the thermodynamic volume can be obtained as
\begin{equation}
V=\left( \frac{\partial M}{\partial P}\right) _{Q,S}=\frac{r_{+}^{3}}{3}%
\left( \frac{r_{+}}{r_{0}}\right) ^{\frac{z}{2}},  \label{V}
\end{equation}%
which is in agreement with $\int A\;dr=\int 4S\;dr=\frac{r_{+}^{3}}{3}$ for
vanishing $z$.

After some calculations, we find that the conserved and
thermodynamic quantities satisfy the following Smarr relation
\begin{equation}
\left( \frac{z}{2}+1\right) M=2TS+\alpha UQ-2PV-\beta V_{0}\Delta _{0},
\label{Smarr}
\end{equation}%
where
\begin{eqnarray*}
\alpha &=&1+\frac{\lambda \sqrt{2z}}{2}, \\
\beta &=&2+\gamma \sqrt{2z},
\end{eqnarray*}%
and $\Delta _{0}$ is a conjugate quantity of $V_{0}$ which is defined as
\begin{equation*}
\Delta _{0}=\left( \frac{\partial M}{\partial V_{0}}\right) _{Q,S,P}=\frac{%
r_{+}^{3}}{8\pi \Gamma }\left( \frac{r_{+}}{r_{0}}\right) ^{\gamma \sqrt{2z}%
+z/2}.
\end{equation*}%
It is worth mentioning that for vanishing $z$, Eq. (\ref{Smarr})
reduces to the known Smarr relation $M=2TS+UQ-2P_{new}V$
($P_{new}=P+\frac{V_{0}}{16\pi }$), as we expected.

Considering Eqs. (\ref{Temp}) and (\ref{P0}), one can find the following
equation of state%
\begin{equation}
P=\frac{3}{4\pi r_{+}(6+z)}\left[ \frac{4\pi T}{\left( \frac{r_{+}}{r_{0}}%
\right) ^{\frac{z}{2}}}-\frac{k}{r_{+}}+\frac{\left( \frac{r_{0}}{r_{+}}%
\right) ^{\lambda \sqrt{2z}}q^{2}}{r_{+}^{3}}-\frac{V_{0}r_{+}}{2}\left(
\frac{r_{+}}{r_{0}}\right) ^{\gamma \sqrt{2z}}\right] ,  \label{P}
\end{equation}%
where $r_{+}$ is a function of thermodynamic volume as indicated in Eq. (\ref%
{V}).

To investigate the existence of van der Waals like phase transition and
critical behavior of the black hole, one can use the unique property of
critical isothermal diagram in the $P-r_{+}$ plot, the inflection point,
\begin{equation}
\left( \frac{\partial P}{\partial r_{+}}\right) _{T}=\left( \frac{\partial
^{2}P}{\partial r_{+}^{2}}\right) _{T}=0.  \label{infel}
\end{equation}

%%%%%%%%%%%%%%%%%%%%%%%%%%%%%%%%%%%%%%%%%%%%%%%%%%%%%
\begin{figure}[tbp]
$%
\begin{array}{ccc}
\epsfxsize=5.5cm \epsffile{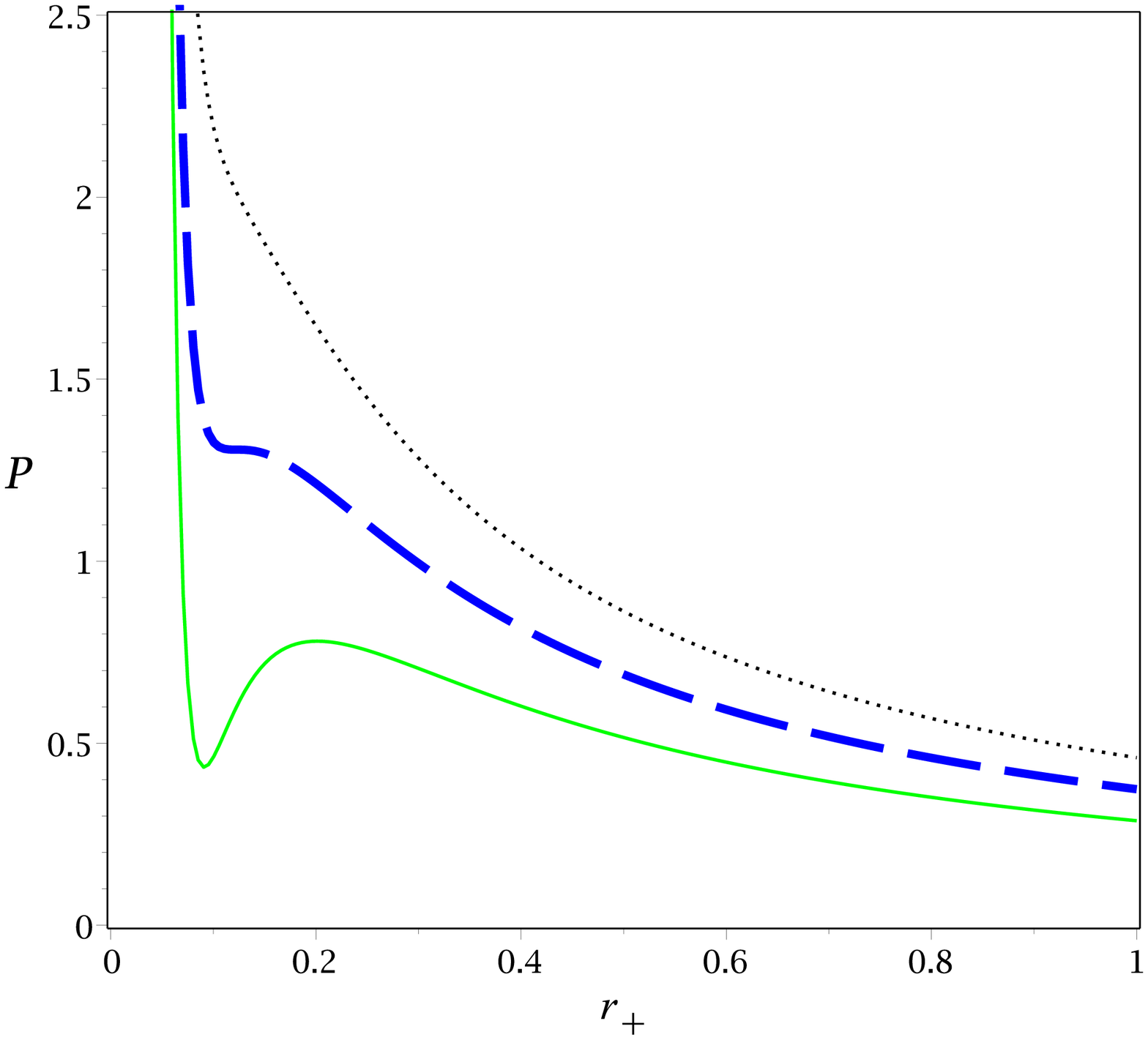} & \epsfxsize=5.5cm %
\epsffile{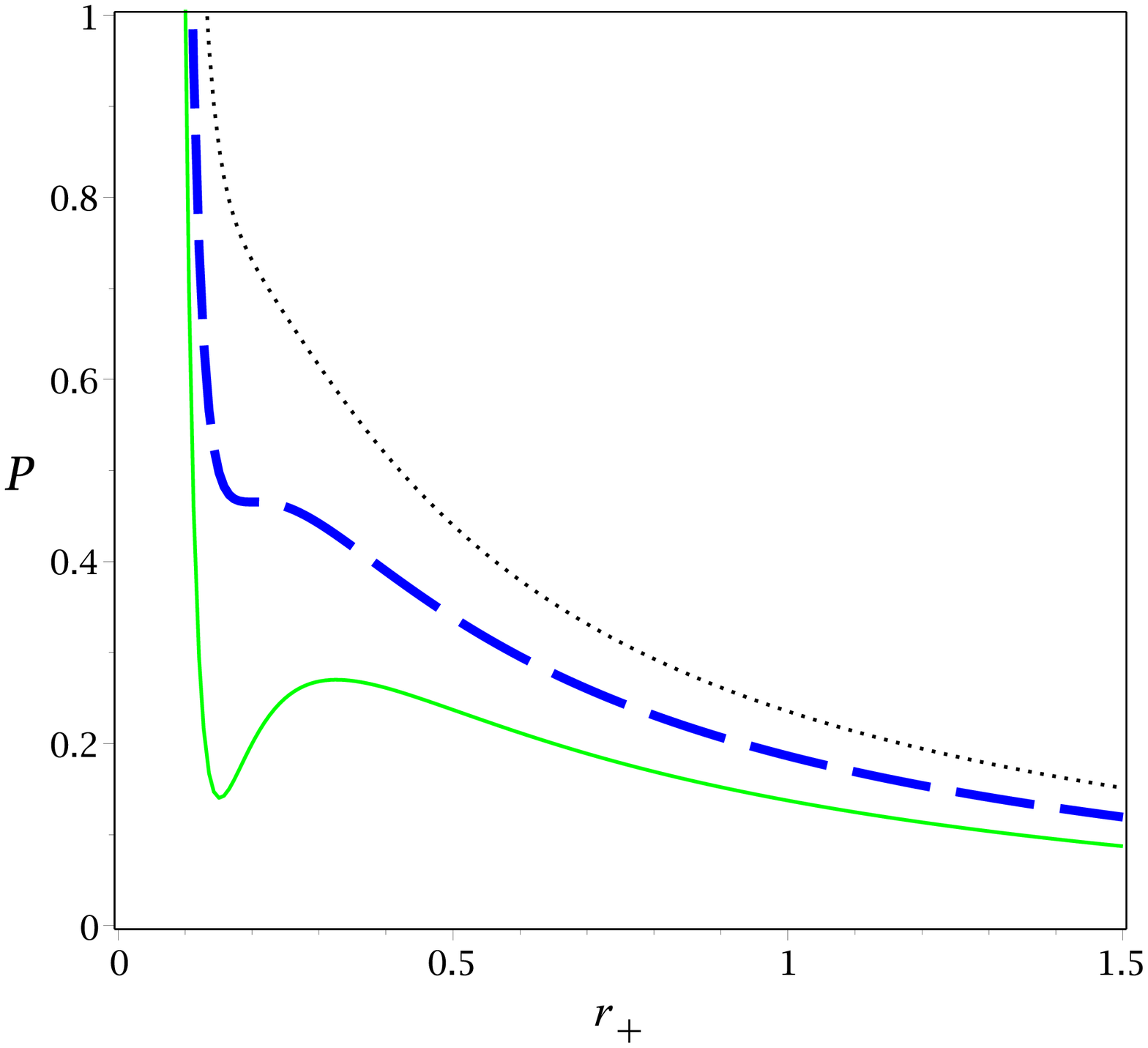} & \epsfxsize=5.5cm \epsffile{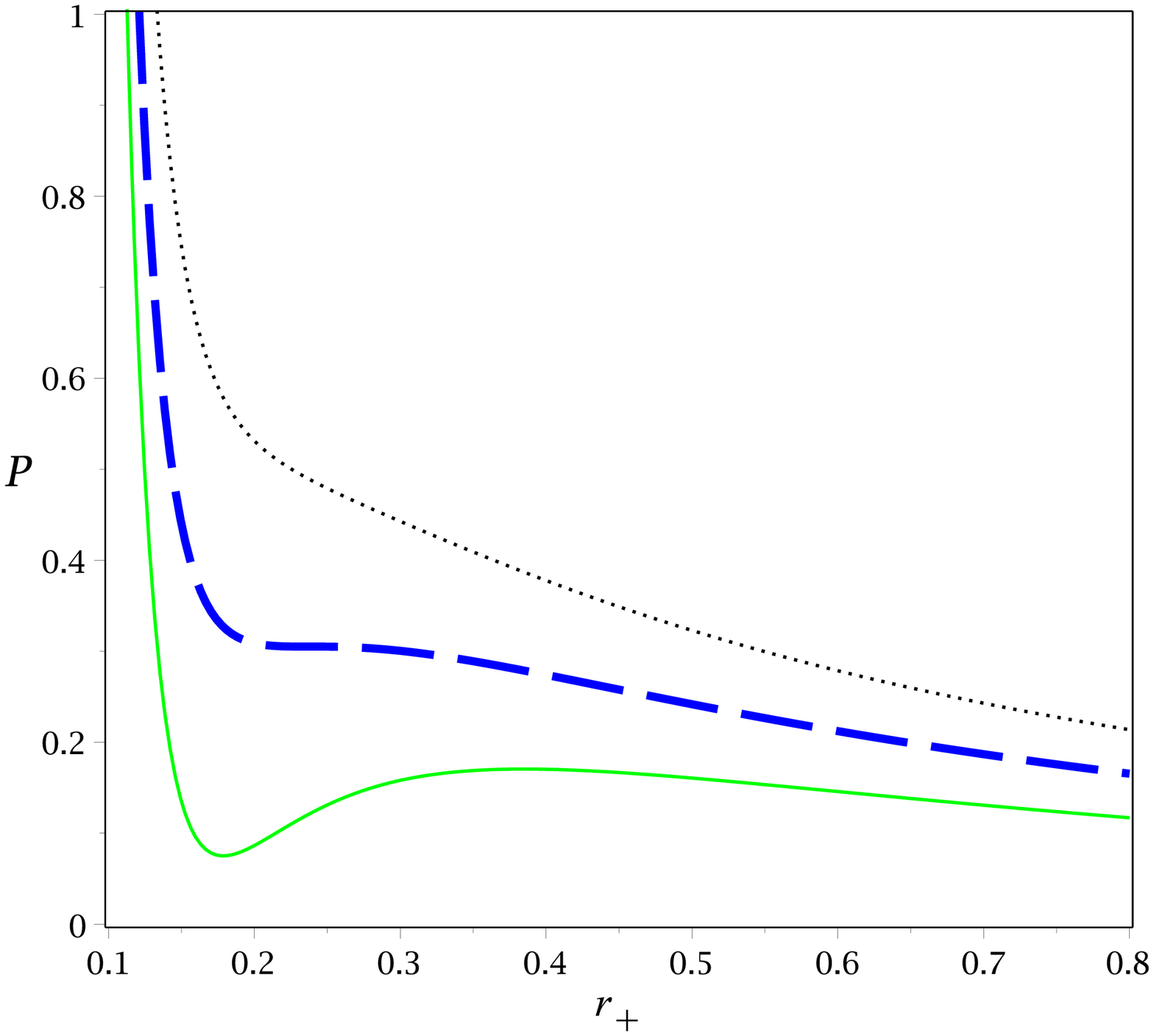}%
\end{array}
$%
\caption{$P$ versus $r_{+}$ for $k=1$, $\protect\lambda=1$, $\protect\gamma%
=1 $, $q=0.05$, $V_{0}=1$, $r_{0}=1$, and $T<T_{c}$ (solid line), $T=T_{c}$
(dashed line) and $T>T_{c}$ (dotted line). \textbf{Left panel:} $z=0$,
\textbf{Middle panel:} $z=0.1$, \textbf{Right panel:} $z=0.2$}
\label{PVz}
\end{figure}
%%%%%%%%%%%%%%%%%%%%%%%%%%%%%%%%%%%%%%%%%%%%%%%%%%%

Traditionary, one should consider Eqs. (\ref{P}) and (\ref{infel})
to obtain the three unknown critical quantities ($r_{c},$ $T_{c}$
and $P_{c}$). Since we cannot obtain the analytical solutions of
Eqs. (\ref{P}) and (\ref{infel}), we continue with numerical
analysis and give suitable tables and figures.

In Fig. \ref{PVz}, we show the van der Waals like behavior of the system. In
order to investigate the effects of various parameters on the critical
behavior, one can study the following tables. According to these tables, we
find that increasing the Lifshitz parameter (decreasing the electric charge)
leads to increasing the critical temperature and pressure and decreasing the
critical horizon. In addition, we find that in order to have critical
behavior or van der Waals like phase transition, one has to imply some
restrictions on the range of free parameters.

%%%%%%%%%%%%%%%%%%%%%%%%%%%%%%%%%%%%%%%%%%%%%%%%%%%%%%%%%%%%%%%%%%%%%%%%%%%%%%%%

\begin{center}
\begin{tabular}{|c|c|c|}
\hline
\begin{tabular}{cccc}
\hline\hline
\hspace{0.4cm}$z$\hspace{0.4cm} & \hspace{0.4cm}$r_{c}$ \hspace{0.4cm} &
\hspace{0.4cm} $T_{c}$\hspace{0.4cm} & \hspace{0.4cm} $P_{c}$\hspace{0.4cm}
\\ \hline\hline
$0.00$ & $0.245$ & $0.433$ & $0.312$ \\ \hline
$0.01$ & $0.279$ & $0.383$ & $0.241$ \\ \hline
$0.05$ & $0.322$ & $0.326$ & $0.175$ \\ \hline
$0.10$ & $0.353$ & $0.288$ & $0.137$ \\ \hline
$0.50$ & $0.486$ & $0.165$ & $0.037$ \\ \hline
$1.00$ & $0.584$ & $0.107$ & $0.001$ \\ \hline
$z\geq 1.1$ & $anomaly$ & $anomaly$ & $negative$ \\ \hline
\end{tabular}
& \ \ \ \ \ \  &
\begin{tabular}{cccc}
\hline\hline
\hspace{0.4cm}$q$\hspace{0.4cm} & \hspace{0.4cm}$r_{c}$ \hspace{0.4cm} &
\hspace{0.4cm} $T_{c}$\hspace{0.4cm} & \hspace{0.4cm} $P_{c}$\hspace{0.4cm}
\\ \hline\hline
$0.001$ & $0.008$ & $10.42$ & $286.64$ \\ \hline
$0.005$ & $0.031$ & $2.99$ & $20.64$ \\ \hline
$0.01$ & $0.054$ & $1.74$ & $6.64$ \\ \hline
$0.05$ & $0.201$ & $0.498$ & $0.465$ \\ \hline
$0.10$ & $0.353$ & $0.288$ & $0.137$ \\ \hline
$0.20$ & $0.613$ & $0.162$ & $0.027$ \\ \hline
$q\geq 0.3$ & $anomaly$ & $anomaly$ & $negative$ \\ \hline
\end{tabular}
\\ \hline
\end{tabular}
\\[0pt]
Left table: $k=1$, $\lambda =1$, $\gamma =1$, $V_{0}=1$, $r_{0}=1$ and $%
q=0.1 $.\\[0pt]
Right table: $k=1$, $\lambda =1$, $\gamma =1$, $V_{0}=1$, $r_{0}=1$ and $%
z=0.1$.
\end{center}

%%%%%%%%%%%%%%%%%%%%%%%%%%%%%%%%%%%%%%%%%%%%%%%%%%%%%%%%%%%%%%%%%%%%%%

Another important quantity to investigate the possible phase
transition (and also its order) is Gibbs free energy. In order to
investigate the local thermodynamic stability one examine the
positivity of the heat capacity while the equilibrium state of a
system is corresponding to the global minimum of the Gibbs free
energy. Since we shall study the thermodynamic behavior of the
system in the canonical (fixed electric charge) ensemble, we can
calculate the Gibbs free energy via $G=M-TS$ with the following
explicit
form%
\begin{equation}
G=\frac{r_{+}}{16\pi }\left( \frac{r_{+}}{r_{0}}\right) ^{\frac{z}{2}}\left[
\frac{(2-z)k}{2+z}+\frac{(4-\Gamma )V_{0}r_{+}^{2}}{2\Gamma \left( \frac{%
r_{0}}{r_{+}}\right) ^{\gamma \sqrt{2z}}}+\frac{q^{2}(\Xi +4)}{\Xi
r_{+}^{2}\left( \frac{r_{+}}{r_{0}}\right) ^{\lambda \sqrt{2z}}}-\frac{4\pi
(z+2)P}{3r_{+}^{-2}}\right] .  \label{Gibbs}
\end{equation}%
The behaviour of Gibbs free energy is essential for finding possible
thermodynamic phase transition and its order. However, it is often
impossible to write, explicitly, the Gibbs free energy as a function of $T$,
$P$ and $Q$ as $G=G(T,P,Q)$, and therefore, we have to use the implicit
numerical calculation to plot $G-T$ diagram.

%%%%%%%%%%%%%%%%%%%%%%%%%%%%%%%%%%%%%%%%%%%%%%%%%%%%%
\begin{figure}[tbp]
$%
\begin{array}{ccc}
\epsfxsize=5.5cm \epsffile{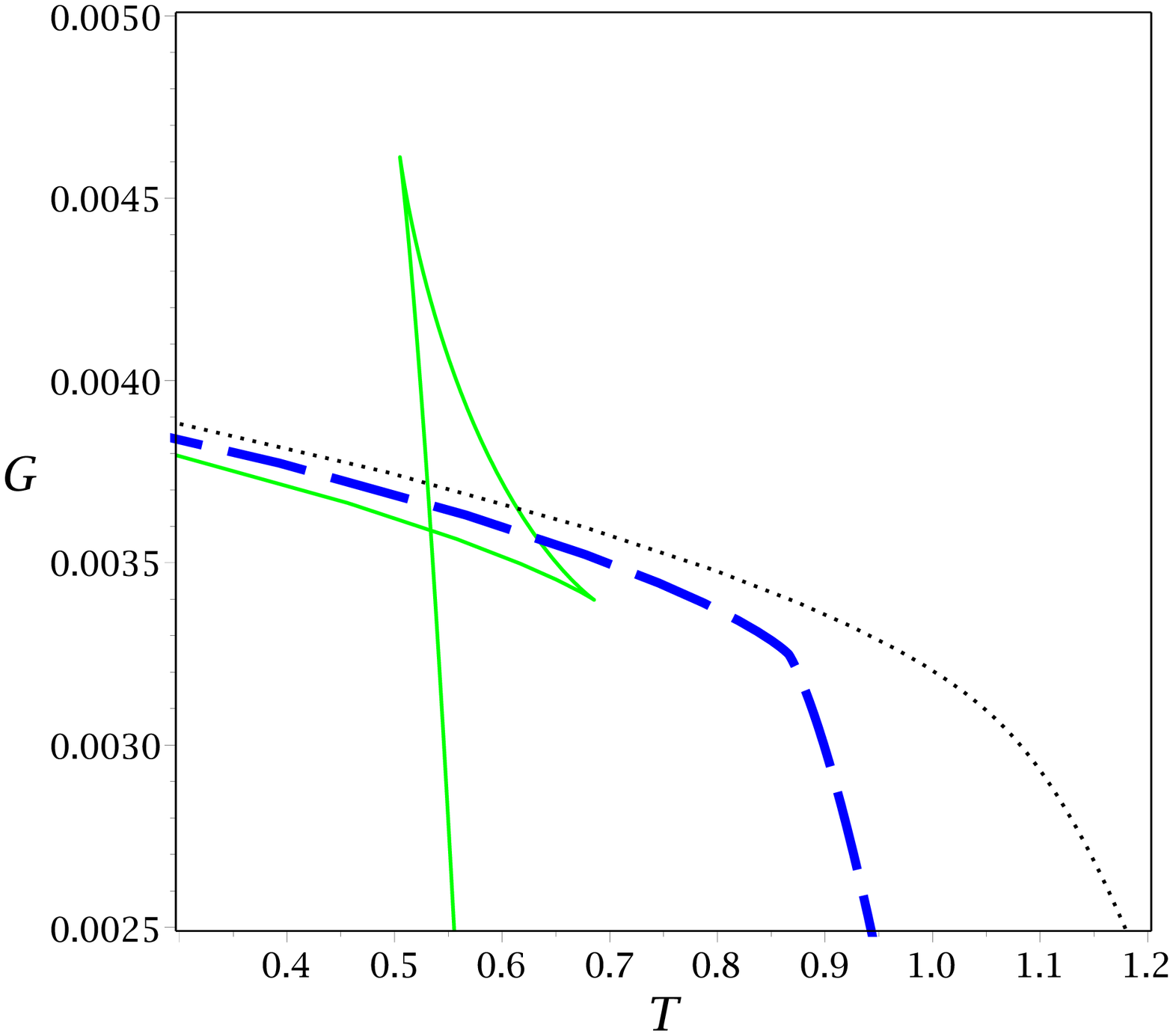} & \epsfxsize=5.5cm %
\epsffile{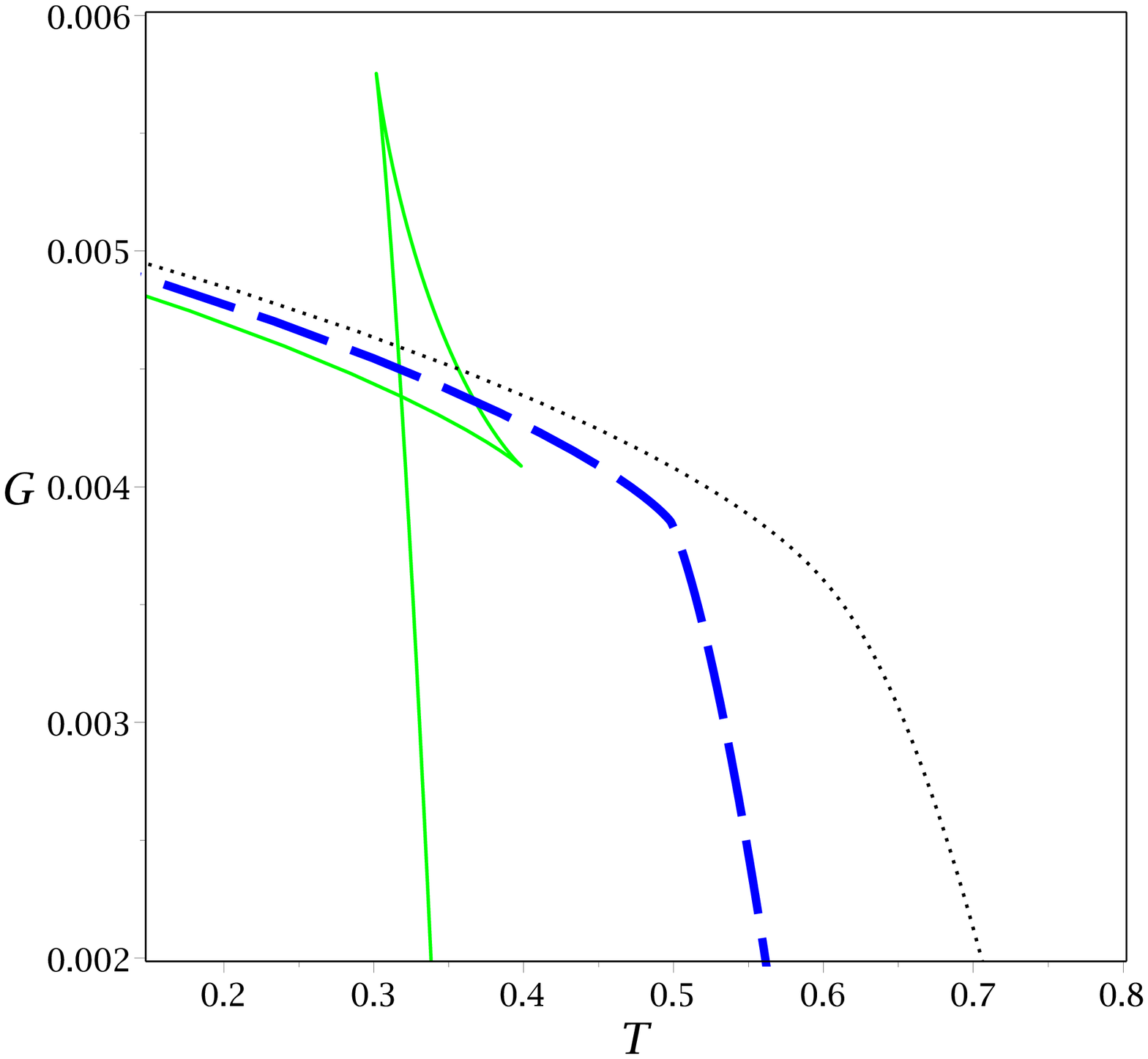} & \epsfxsize=5.5cm \epsffile{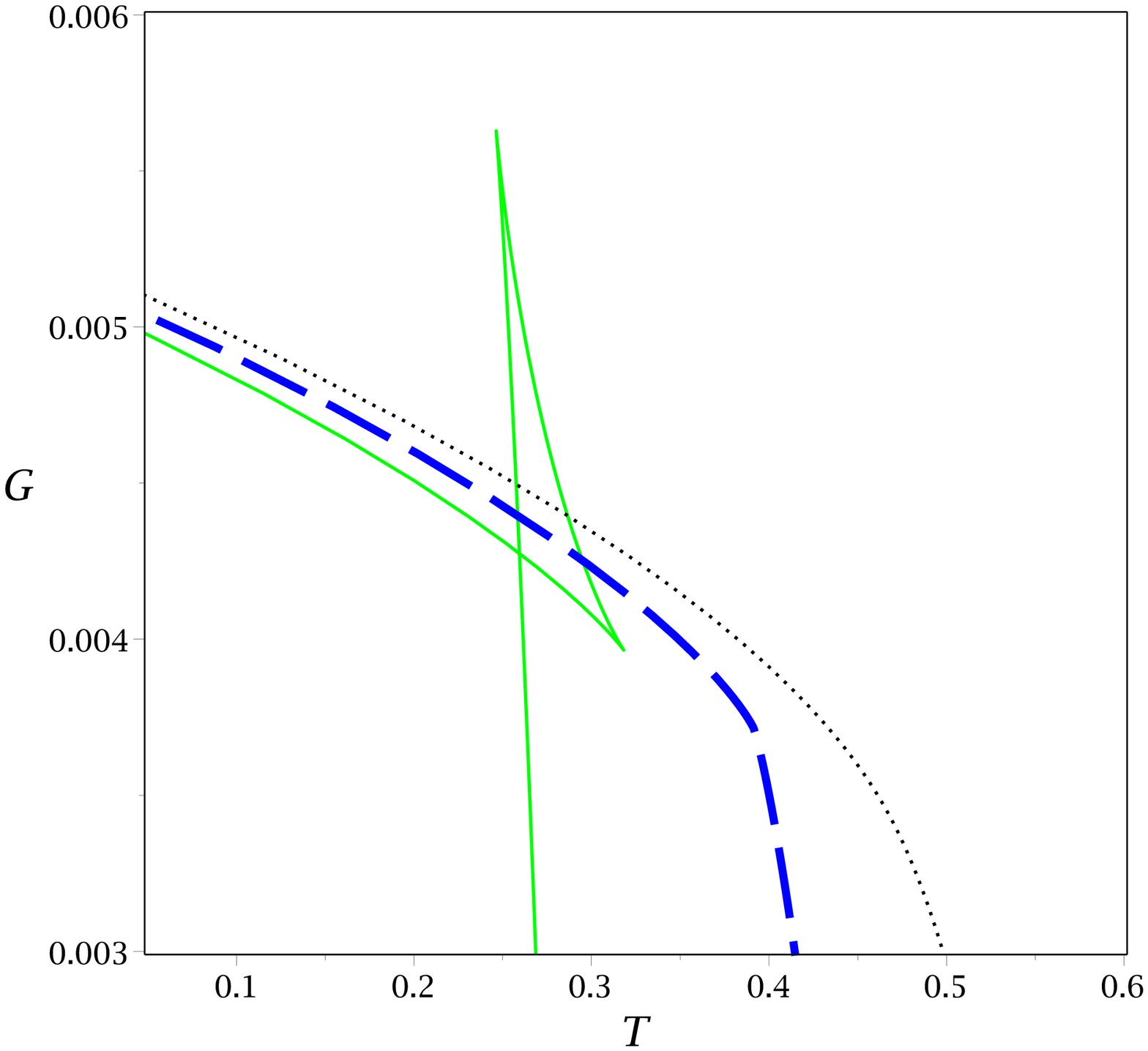}%
\end{array}
$%
\caption{Gibbs free energy versus temperature for $k=1$, $\protect\lambda=1$%
, $\protect\gamma=1 $, $q=0.05$, $V_{0}=1$, $r_{0}=1$, and $P<P_{c}$ (solid
line), $P=P_{c}$ (dashed line) and $P>P_{c}$ (dotted line). \textbf{Left
panel:} $z=0$, \textbf{Middle panel:} $z=0.1$, \textbf{Right panel:} $z=0.2$}
\label{GTz}
\end{figure}
%%%%%%%%%%%%%%%%%%%%%%%%%%%%%%%%%%%%%%%%%%%%%%%%%%%

The behavior of Gibbs free energy with respect to temperature may be
investigated via plotting the graph of isobaric $G-T$. Regarding Fig. \ref%
{GTz}, one finds the characteristics swallow-tail behavior of $G-T$ diagrams
guarantees the existence of a first order phase transition.

As a final comment of this section, we should note that by adjusting the
free parameters of the system, we can find a van der Waals like behavior for
planar and hyperbolic topologies. For example, we can set $z=2$, $\gamma
=2\lambda =2r_{0}=-1$ and $V_{0}=24$ for $k=-1$ (and $V_{0}=16$ for $k=0$)
to obtain
\begin{equation}
B(r)=1-\frac{m}{r^{2}}+\frac{2q^{2}}{r}-\frac{\Lambda r^{2}}{4},
\label{Sol-2}
\end{equation}%
which is nothing but a Reissner-Nordstr\"{o}m-AdS black hole in which charge
$"q^{2}"$ and mass $"-m"$ change their roles, and therefore, the possibility
of a van der Waals like phase transition is evident.

\subsection{Heat Capacity and Thermal Stability}

Here, we will explain the criteria of having thermal stability of
black holes. Analyzing the local thermal stability of a black hole
can be performed in the canonical ensemble. In the context of
extended phase space thermodynamics, the pressure and electric
charge are kept constant in the canonical ensemble, and therefore,
the stability criterion will be the positivity of heat capacity,
i.e.,
\begin{equation}
C_{P,Q}=T\left( \frac{\partial S}{\partial T}\right) _{P,Q}=\frac{M_{S}}{%
M_{SS}},  \label{CQ1}
\end{equation}%
where $M_{S}=\left( \frac{\partial M}{\partial S}\right) _{P,Q}$ and $%
M_{SS}=\left( \frac{\partial ^{2}M}{\partial S^{2}}\right) _{P,Q}$. After
some calculations and simplification, one finds
\begin{equation}
C_{P,Q}=\frac{V_{0}r_{+}^{4}\left( \frac{r_{+}}{r_{0}}\right) ^{\gamma \sqrt{%
2z}}-2q^{2}\left( \frac{r_{0}}{r_{+}}\right) ^{\lambda \sqrt{2z}}+\frac{%
2r_{+}^{2}\left[ 3k+4\pi (z+6)Pr_{+}^{2}\right] }{3}}{V_{0}r_{+}^{2}(\Gamma
-4)\left( \frac{r_{+}}{r_{0}}\right) ^{\gamma \sqrt{2z}}+\frac{2(\Xi +4)q^{2}%
}{r_{+}^{2}}\left( \frac{r_{0}}{r_{+}}\right) ^{\lambda \sqrt{2z}}+\frac{%
2(z-2)\left[ 3k+\frac{4\pi (z+6)(z+2)}{(z-2)}Pr_{+}^{2}\right] }{3}}.
\end{equation}

%%%%%%%%%%%%%%%%%%%%%%%%%%%%%%%%%%%%%%%%%%%%%%%%%%%%%
\begin{figure}[tbp]
$%
\begin{array}{ccc}
\epsfxsize=5.8cm \epsffile{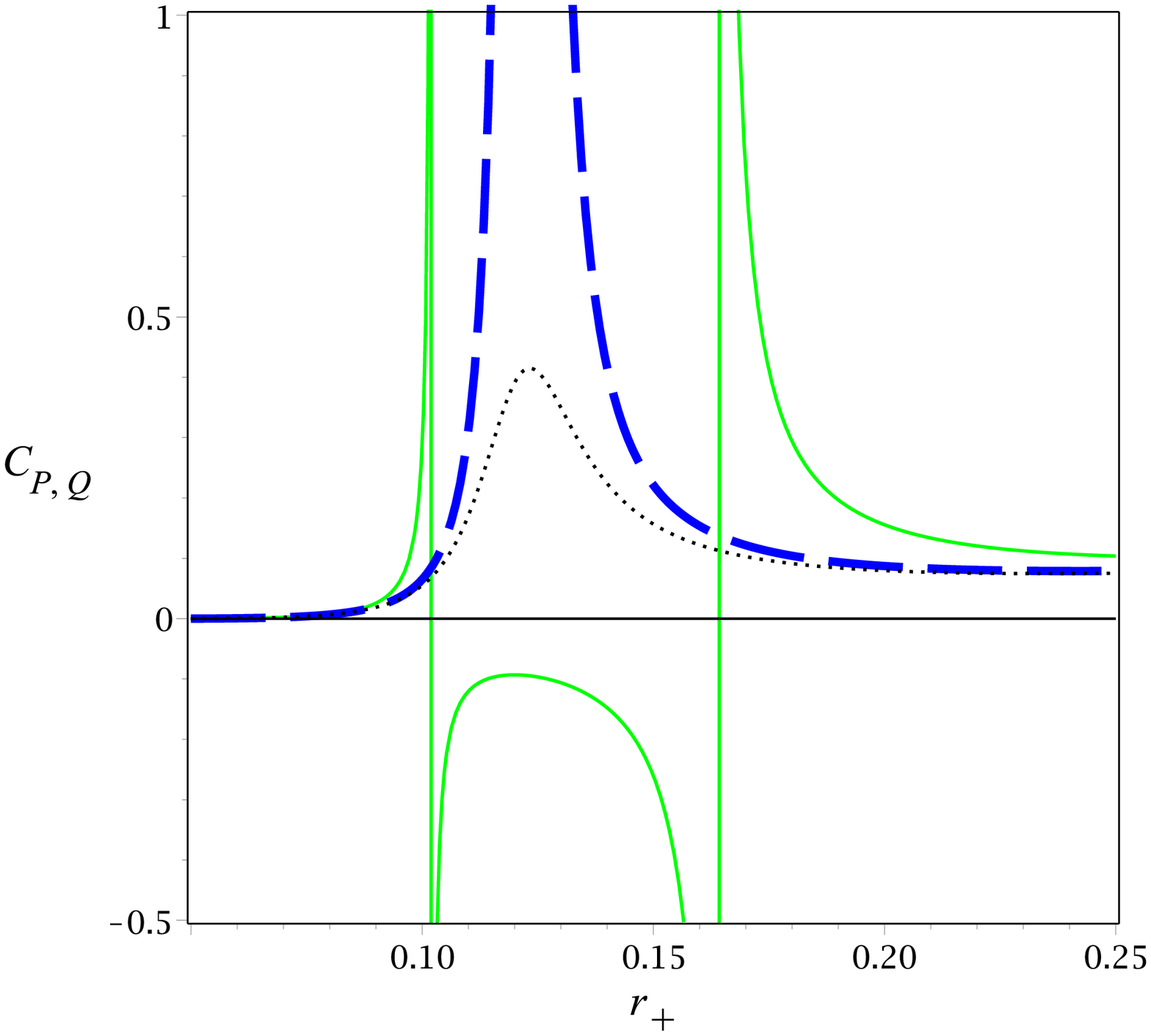} & \epsfxsize=5.4cm %
\epsffile{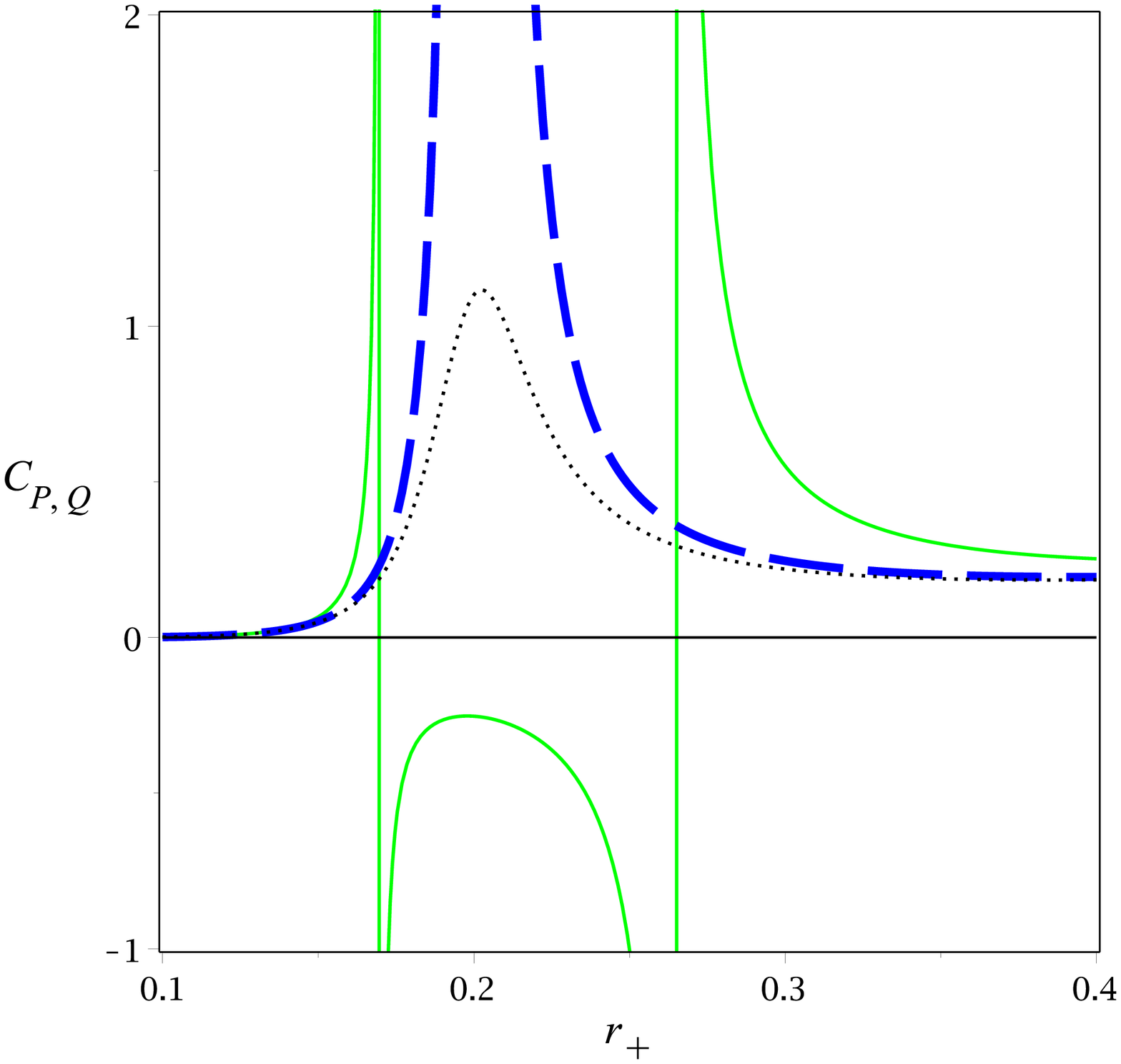} & \epsfxsize=6cm \epsffile{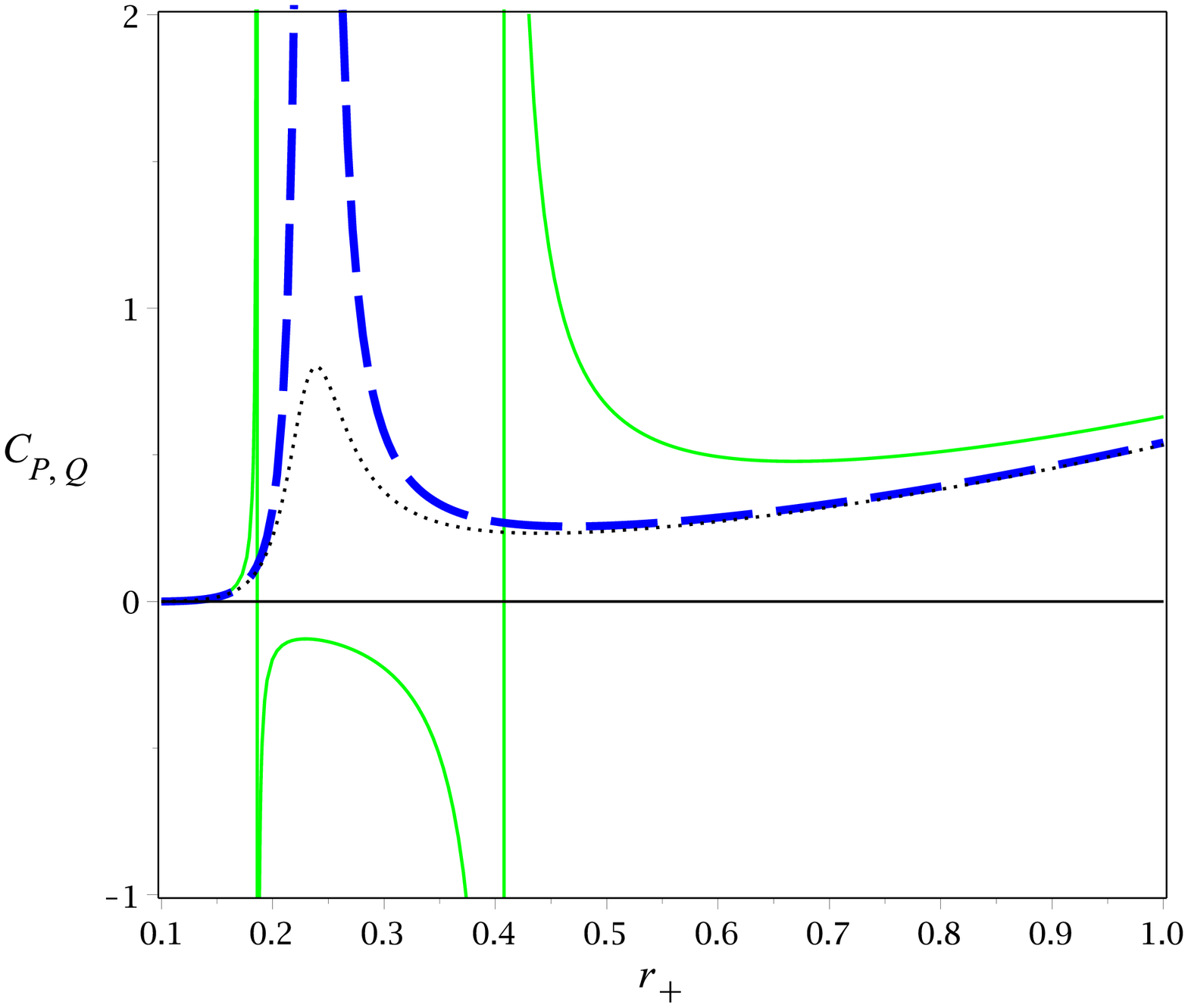}%
\end{array}
$%
\caption{Heat capacity versus $r_{+}$ for $k=1$, $\protect\lambda=1$, $%
\protect\gamma=1 $, $q=0.05$, $V_{0}=1$, $r_{0}=1$, and $P<P_{c}$ (solid
line), $P=P_{c}$ (bold line) and $P>P_{c}$ (dotted line). \textbf{Left panel:%
} $z=0$, \textbf{Middle panel:} $z=0.1$, \textbf{Right panel:} $z=0.2$}
\label{CQz}
\end{figure}
%%%%%%%%%%%%%%%%%%%%%%%%%%%%%%%%%%%%%%%%%%%%%%%%%%%

Regarding the heat capacity, it is notable that its divergence points may be
interpreted as phase transitions of our system. In order to have a van der
Waals like behavior, we should obtain two divergence points for the heat
capacity with $C_{P,Q}<0$ between them. Figure \ref{CQz} is plotted to
confirm the existence of a first order phase transition. We find that for $%
P<P_{c}$, there are two divergence points related to a first order phase
transition. For $P\rightarrow P_{c}$, the two divergence points emerge to
one and negative region disappears at $P=P_{c}$ which is related to the
critical point. As expected, for $P>P_{c}$, there is no divergency for the
heat capacity and black holes are thermally stable.

\subsection{Response functions}

One of the interesting methods including the fundamental
thermodynamic information is extracting the response functions of
the system under consideration. One of these response functions is
the heat capacity which is related to the second derivatives of
the Gibbs free energy with respect to the temperature, as
$C_{P,Q}=T\left( \frac{\partial S}{\partial T}\right)
_{P,Q}=\frac{T}{\left( \frac{\partial ^{2}M}{\partial S^{2}}\right) _{P,Q}}%
=-T\left( \frac{\partial ^{2}G}{\partial T^{2}}\right) _{P,Q}$.
Other (Gibbsian) response functions are related to the second
derivatives of the Gibbs free energy with respect to the
independent thermodynamic quantities. Regarding the
functional form of the Gibbs free energy $G=G(T,Q,P)$ given in Eq. (\ref%
{Gibbs}), one finds that the response functions can be related to $\left(
\frac{\partial ^{2}G}{\partial P^{2}}\right) _{T,Q}$, $\left( \frac{\partial
^{2}G}{\partial Q^{2}}\right) _{T,P}$, $\left( \frac{\partial ^{2}G}{%
\partial T\partial P}\right) _{Q}$, $\left( \frac{\partial ^{2}G}{\partial
T\partial Q}\right) _{P}$ and $\left( \frac{\partial
^{2}G}{\partial P\partial Q}\right) _{T}$. Since we look for the
possible van der Waals like phase transition, in this paper, we
focused on the heat capacity and put other response functions
aside. However, one can investigate divergence points of other
response functions and their relation to novel phase transitions.

\subsection{Geometrical thermodynamics}

Regarding the geometrical thermodynamic method, the thermodynamic phase
space and the space of equilibrium state are involved. Many attempts were
made to describe the thermodynamic behavior of black holes, geometrically,
based on the defined metric on the equilibrium space \cite%
{Weinhold,Weinhold01}. In particular, the Weinhold and Ruppeiner
metrics were used in order to link the curvature singularities of
defined thermodynamic metric and heat capacity divergences,
directly, but the results caused contradictory outcomes. There are
many examples that the singularities of these metrics are not
located on the phase transition points, or in some cases, there
are a number of singularities before and after those points. As an
example, for the Kerr black hole the Weinhold metric predicts that
there is no phase transition, while Ruppeiner one predicts that
with a certain thermodynamic potential, the phase transition is in
accordance with the standard results of thermal properties of
black holes \cite{Quevedo1,Quevedo11}.

It is shown that these inconsistencies arise from the fact that both
Weinhold and Ruppeiner metrics are not Legendre invariant, which makes the
metrics not suitable to describe the thermodynamics of the system. Legendre
invariance plays a crucial role in geometrical thermodynamics and it means
when a particular demonstration is chosen for description of a system, the
Legendre transformations in the new form should have the same information as
the initial representation. Therefore, Legendre invariance is one of the
important characteristics of thermodynamic metrics when making the desired
geometry.

In order to resolve such problems, Quevedo and his colleagues proposed a
method to find the metric that is Legendre invariance \cite%
{Quevedo1,Quevedo11}. Although the Quevedo metric solved most of the
problems of the previous metrics, it faced a problem in several specific
cases. To solve its problem, a new metric has been recently introduced \cite%
{HPEM} in which the problem of additional singularities that are not
consistent with any of the phase transition points is not observed.

In what follows, we intend to introduce these thermodynamical metrics and
discuss their results in our case. The first metric which was introduced by
Weinhold is given by

%%%%%%%%%%%%%%%%%%%%%%%%%%%%%%%%%%%%%%%%%%%%%%%%%%%%%
\begin{figure}[tbp]
$%
\begin{array}{ccc}
\epsfxsize=5.5cm \epsffile{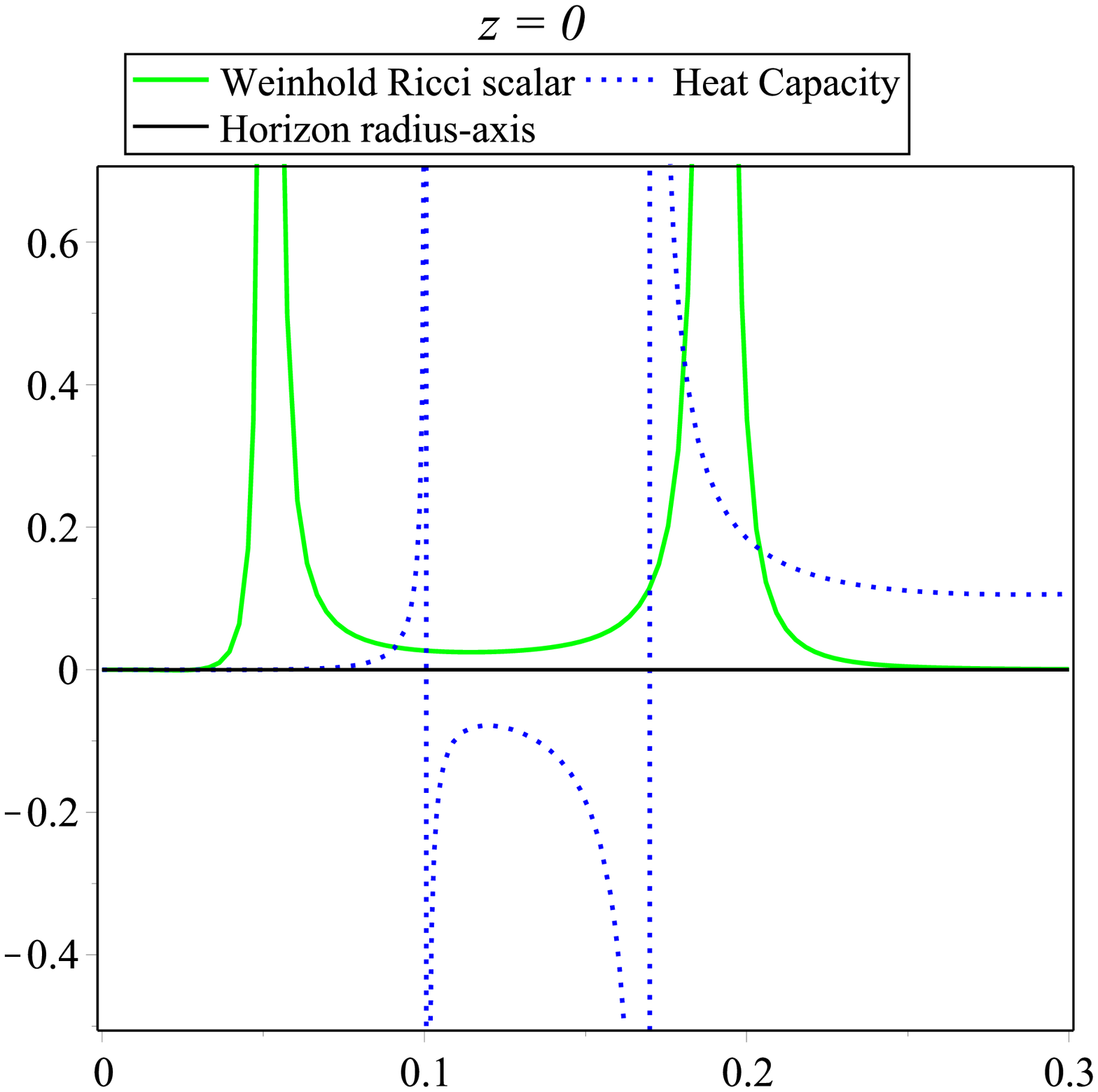} & \epsfxsize=5.5cm %
\epsffile{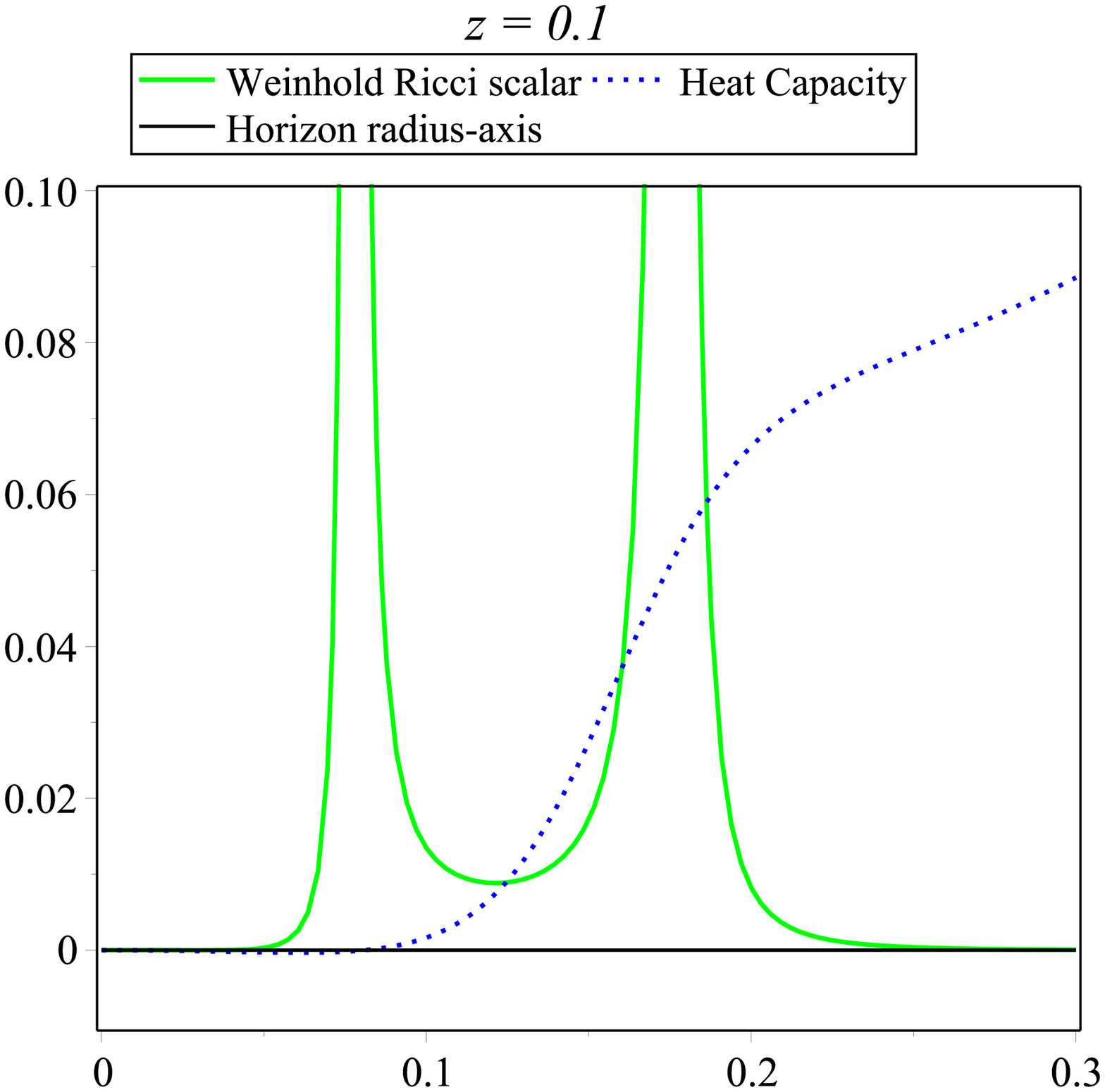} & \epsfxsize=5.5cm \epsffile{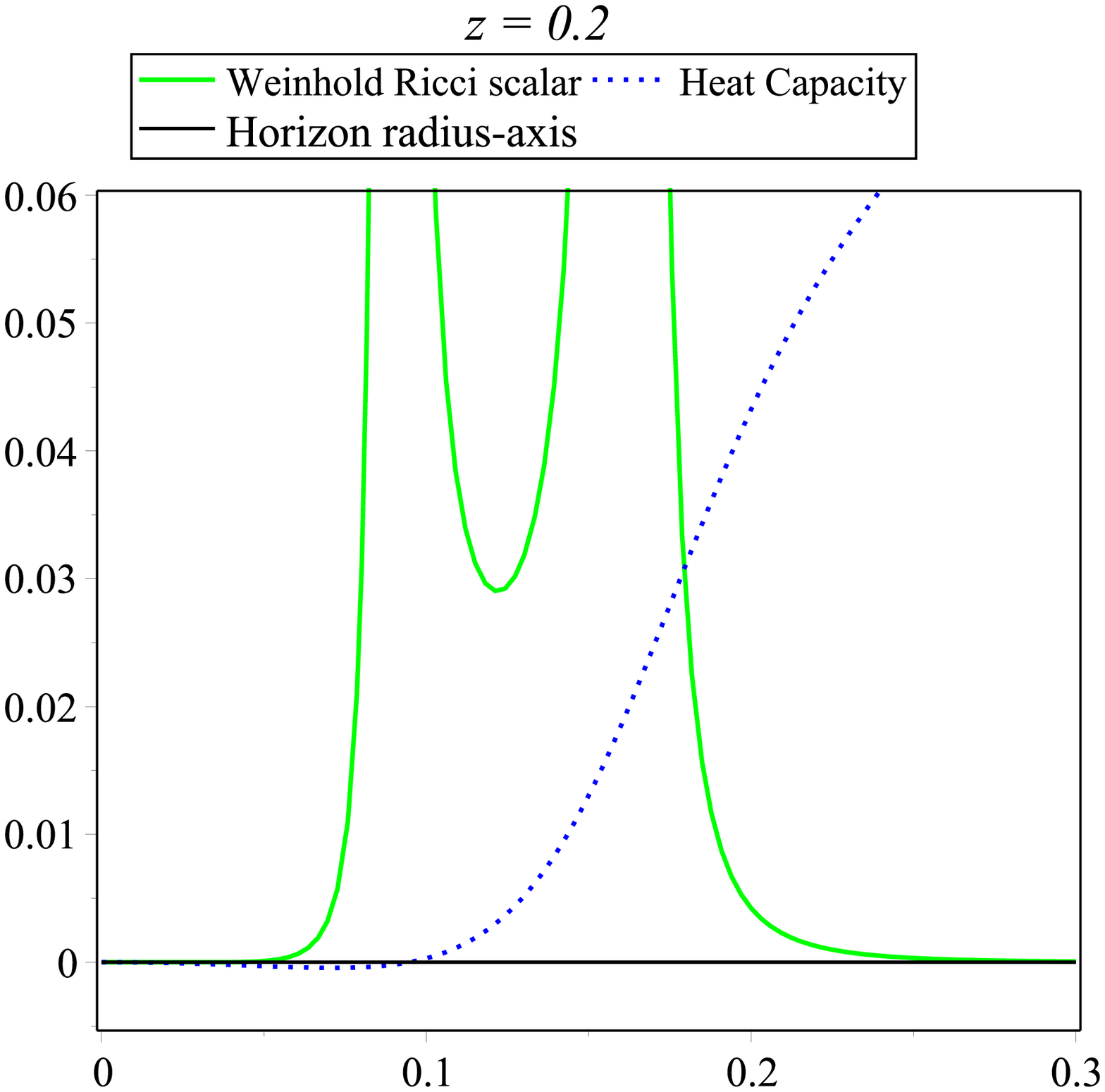}%
\end{array}
$%
\caption{ Heat Capacity (dotted line) and the Ricci scalar of
Weinhold metric
(solid line) versus $r_{+}$ for $k=1$, $\protect\lambda=1$, $\protect\gamma%
=1 $, $q=0.05$, $V_{0}=1$, $r_{0}=1$ and $P<P_{c}$. \textbf{Left panel:} $%
z=0 $, \textbf{Middle panel:} $z=0.1$, \textbf{Right panel:}
$z=0.2$ \emph{\textbf{(For more clarifications we used different
scales).}}} \label{Weinz}
\end{figure}
%%%%%%%%%%%%%%%%%%%%%%%%%%%%%%%%%%%%%%%%%%%%%%%%%%%

\begin{equation}
dS_{W}^{2}=Mg_{ab}^{W}dX^{a}dX^{b},  \label{Weinhold}
\end{equation}%
where $M$ is the mass which is a function of extensive quantities such as
entropy and electric charge and hereafter we use the notation $%
M_{X}=\partial M/\partial X$ and $M_{XX}=\partial ^{2}M/\partial X^{2}$.
Finding divergence points of the Ricci scalar is interesting for us. In
order to find these points, we involve the denominator of the Ricci scalar,
since its numerator is a finite smooth function. The denominator of Weinhold
Ricci scalar for our system is
\begin{equation}
denom(R_{W})=M^{3}\left( M_{SS}M_{QQ}-M_{SQ}^{2}\right) ^{2}.  \label{DW}
\end{equation}

It is clear that there is at least a divergence point for the Ricci scalar
when $M_{SS}M_{QQ}=M_{SQ}^{2}$, which is not consistent with any divergency
of the heat capacity. We plot both the Ricci scalar and heat capacity in
Fig. \ref{Weinz} to inquire into the mentioned behavior. As one can see,
divergence points of the heat capacity are not coincidence with those of the
Ricci scalar of Weinhold metric.
%%%%%%%%%%%%%%%%%%%%%%%%%%%%%%%%%%%%%%%%%%%%%%%%%%%%%
\begin{figure}[tbp]
$%
\begin{array}{ccc}
\epsfxsize=5.5cm \epsffile{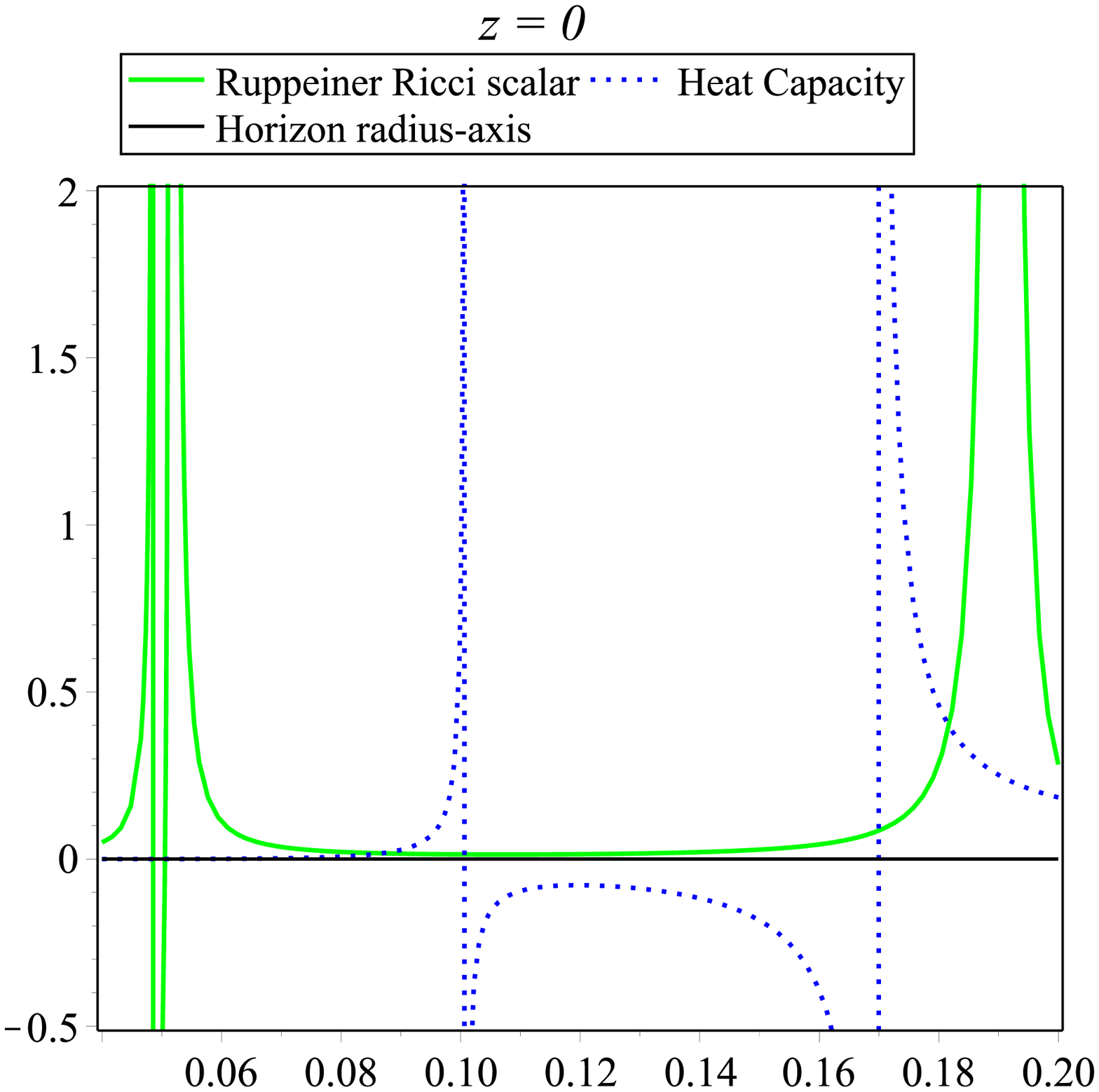} & \epsfxsize=5.5cm %
\epsffile{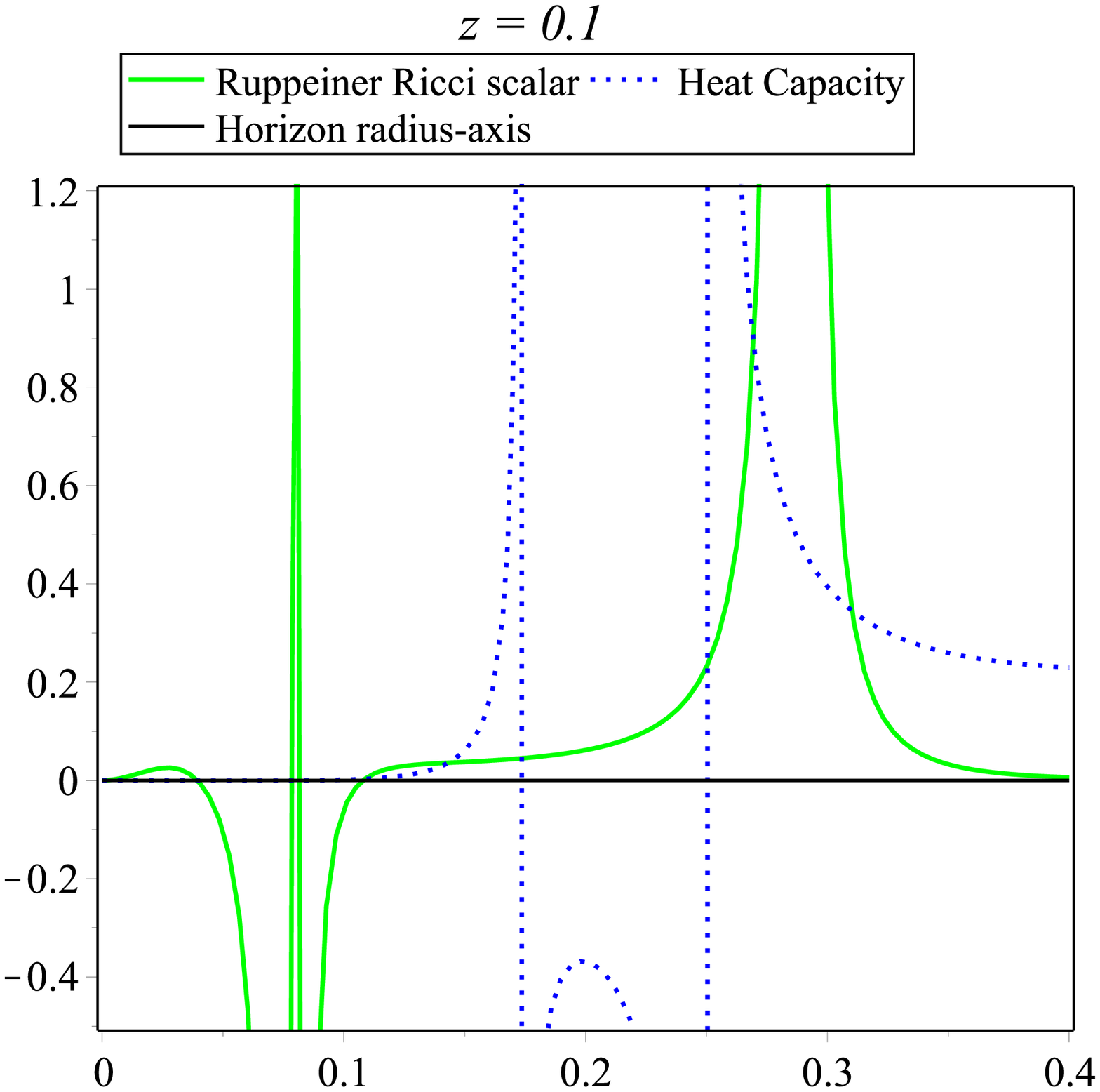} & \epsfxsize=5.5cm \epsffile{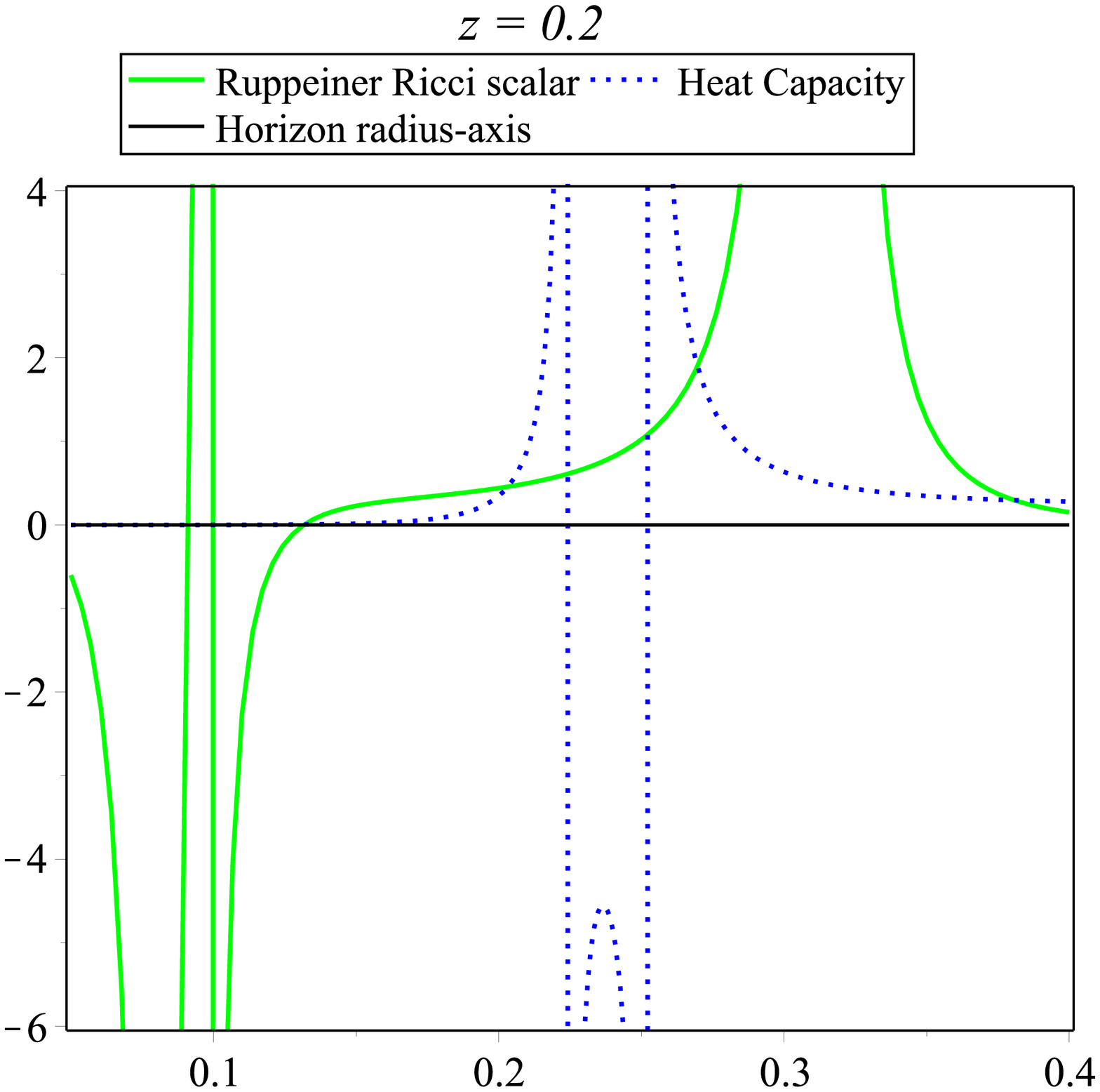}%
\end{array}
$%
\caption{Heat Capacity (dotted line) and the Ricci scalar of
Ruppeiner
metric (solid line) versus $r_{+}$ for $k=1$, $\protect\lambda=1$, $\protect%
\gamma=1 $, $q=0.05$, $V_{0}=1$, $r_{0}=1$ and $P<P_{c}$. \textbf{Left panel:%
} $z=0$, \textbf{Middle panel:} $z=0.1$, \textbf{Right panel:}
$z=0.2$ \emph{\textbf{(For more clarifications we used different
scales).}}} \label{Rupz}
\end{figure}
%%%%%%%%%%%%%%%%%%%%%%%%%%%%%%%%%%%%%%%%%%%%%%%%%%%

The Ruppeiner metric was the second metric which is related to Weinhold
metric by a Legendre transformation as
\begin{equation}
dS_{R}^{2}=-MT^{-1}g_{ab}^{W}dX^{a}dX^{b}.  \label{Ruppeiner}
\end{equation}%
Taking into account the relation between these two metrics, one
can find the denominator of Ruppeiner's Ricci scalar as
\begin{equation}
denom(R_{R})=M^{3}T\left( M_{SS}M_{QQ}-M_{SQ}^{2}\right) ^{2}.  \label{DR}
\end{equation}

It is evident that the same additional divergency of the Weinhold
metric is presented here. In Fig. \ref{Rupz}, we plot the
Ruppeiner Ricci scalar which shows that its divergencies are not
matched with those of the heat capacity, like Weinhold case.

%%%%%%%%%%%%%%%%%%%%%%%%%%%%%%%%%%%%%%%%%%%%%%%%%%%%%
\begin{figure}[tbp]
$%
\begin{array}{ccc}
\epsfxsize=5.5cm \epsffile{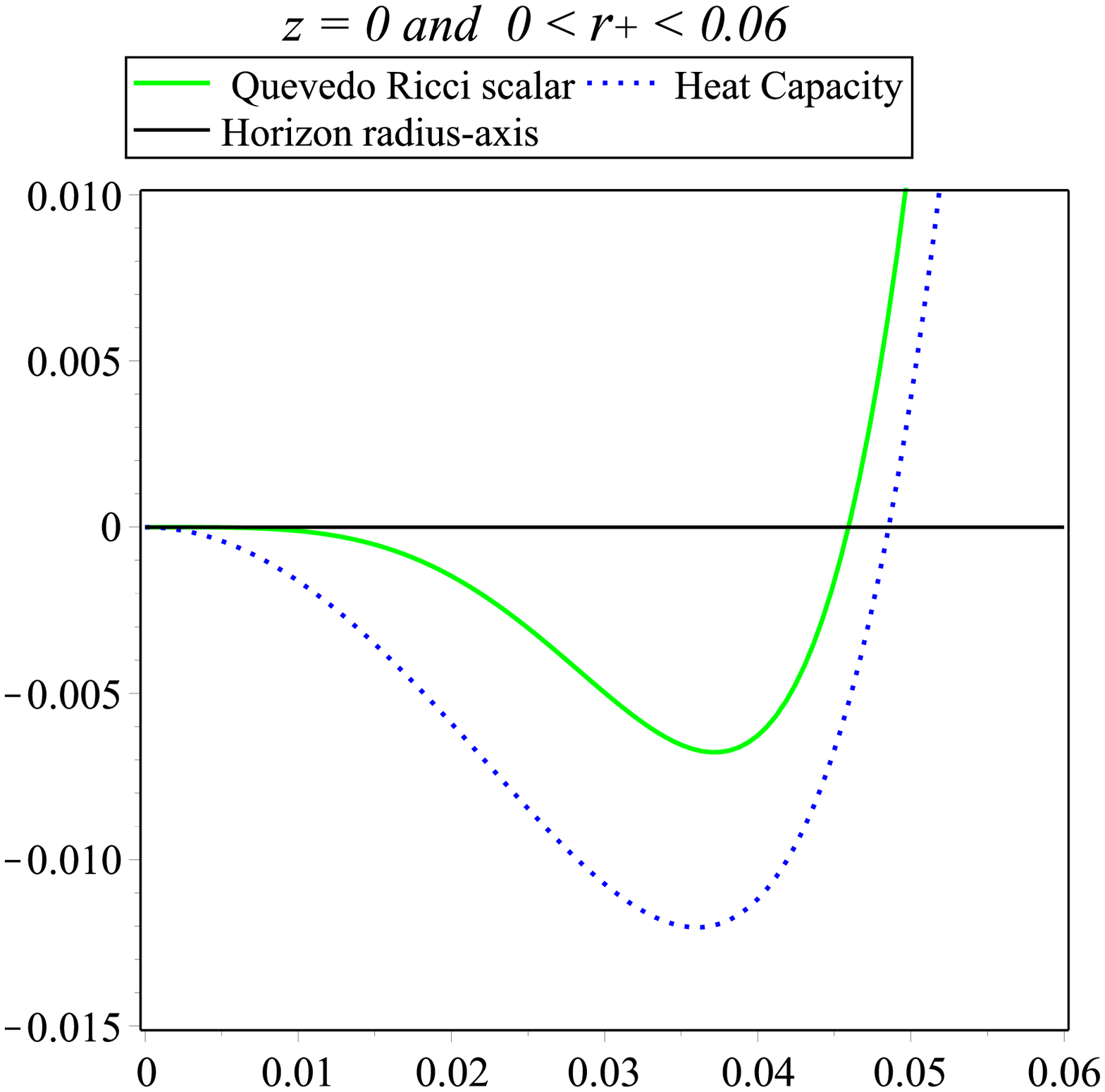} & \epsfxsize=5.5cm %
\epsffile{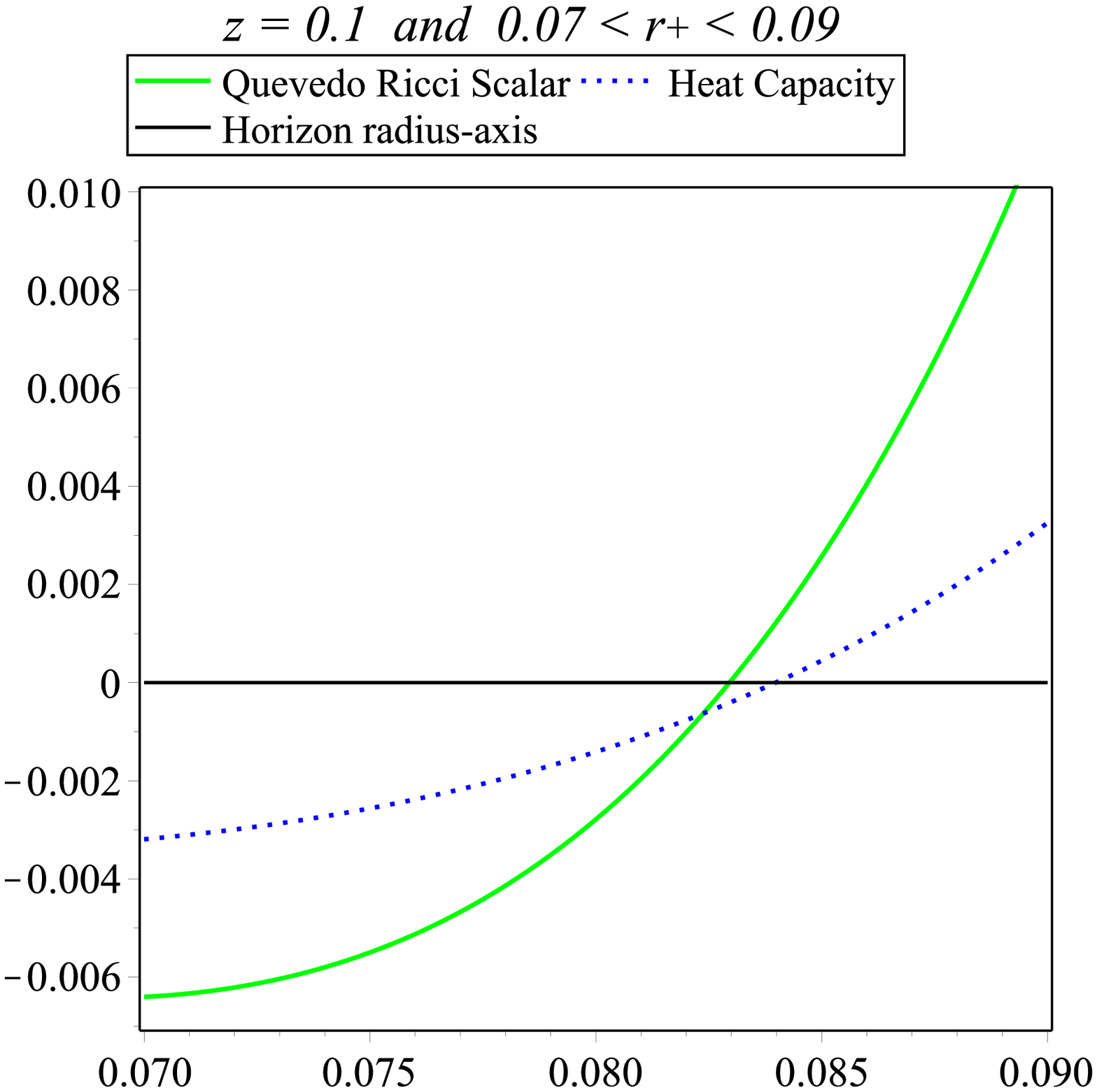} & \epsfxsize=5.5cm \epsffile{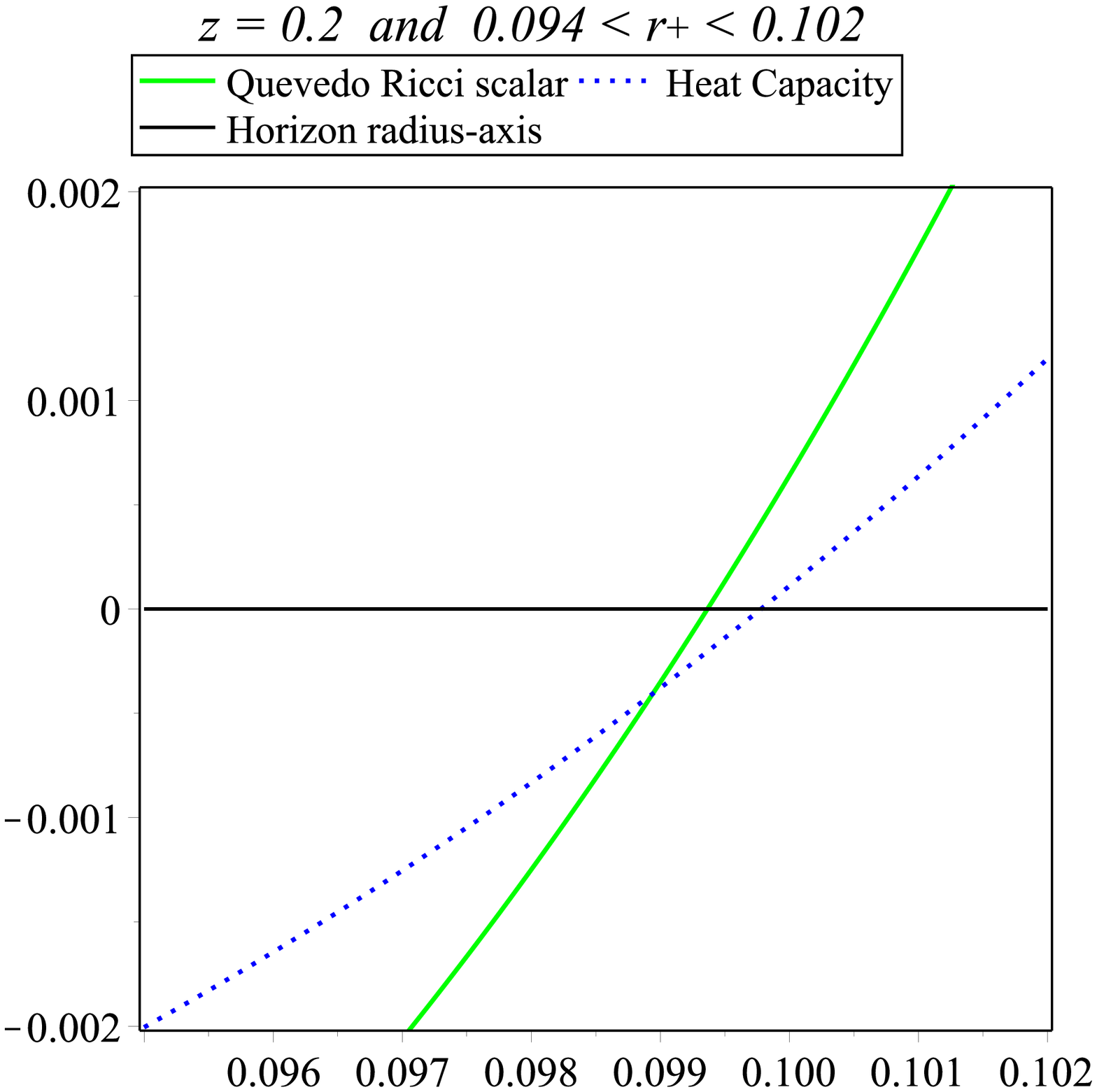} \\
\epsfxsize=5.5cm \epsffile{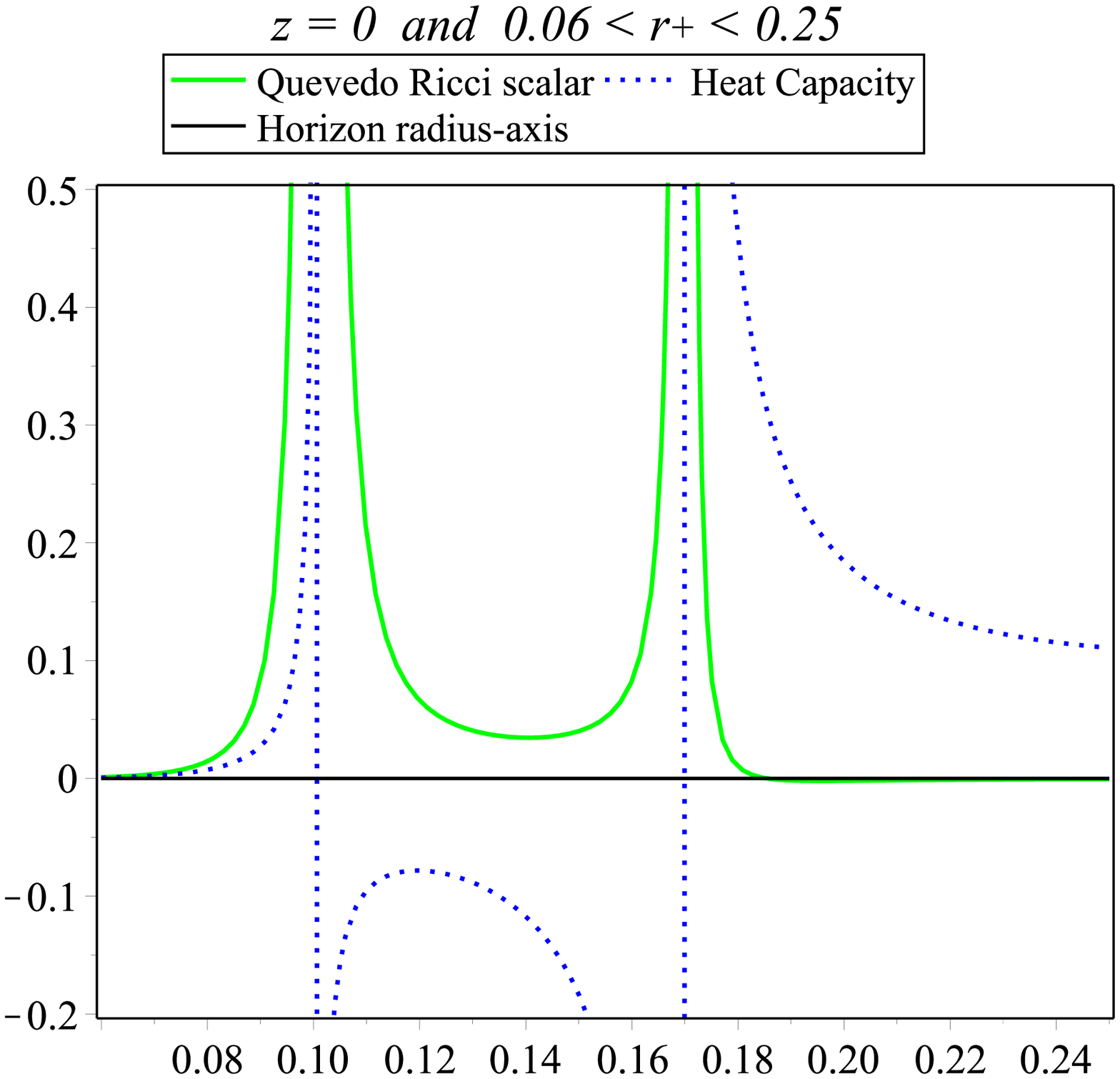} & \epsfxsize=5.5cm %
\epsffile{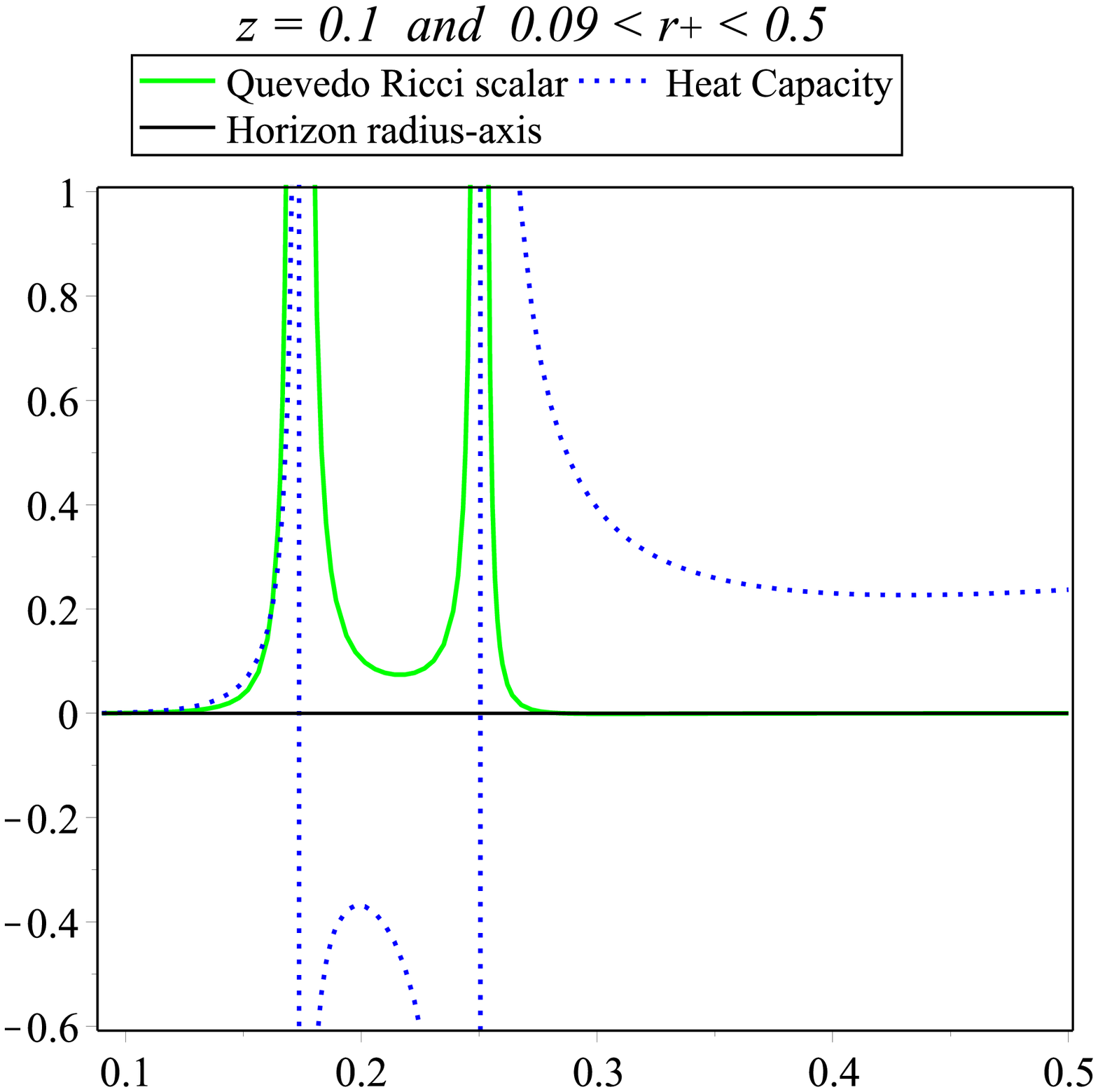} & \epsfxsize=5.5cm \epsffile{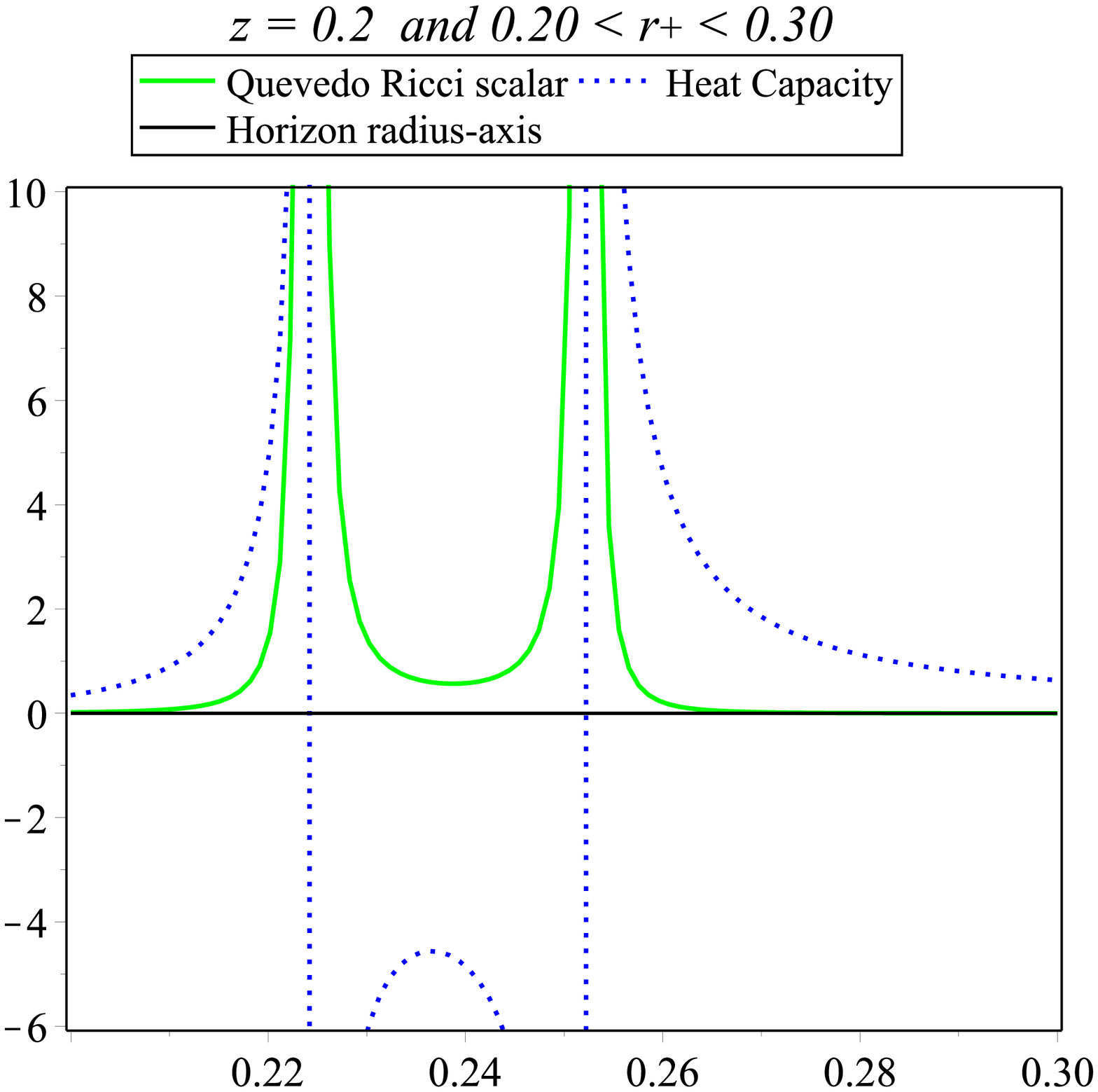}%
\end{array}
$%
\caption{Heat Capacity (dotted line) and the Ricci scalar of
Quevedo metric
(solid line) versus $r_{+}$ for $k=1$, $\protect\lambda=1$, $\protect\gamma%
=1 $, $q=0.05$, $V_{0}=1$, $r_{0}=1$ and $P<P_{c}$. \textbf{Left panels:} $%
z=0 $, \textbf{Middle panels:} $z=0.1$, \textbf{Right panels:}
$z=0.2$ \emph{\textbf{(For more clarifications we used different
scales).}}} \label{Quevz}
\end{figure}
%%%%%%%%%%%%%%%%%%%%%%%%%%%%%%%%%%%%%%%%%%%%%%%%%%%

As these two metrics cannot properly describe our system, we employ Quevedo
metric in order to remove such failures. The Quevedo metric has the
following form
\begin{equation}
dS_{Q}^{2}=\Omega \left( -M_{SS}dS^{2}+M_{QQ}dQ^{2}\right) ,  \label{Quevedo}
\end{equation}%
where $\Omega =SM_{S}+QM_{Q}$. The denominator of Quevedo Ricci scalar is%
\begin{equation}
denom(R_{Q})=M_{SS}^{2}M_{QQ}^{2}\left( SM_{S}+QM_{Q}\right) ^{3}.
\label{DQ}
\end{equation}%
Although vanishing of $M_{SS}$ indicates that such divergency is
the same as that of the heat capacity, $M_{QQ}=0$ and
$SM_{S}+QM_{Q}=0$ deal with additional (inconsistent)
divergencies. We can find the mentioned behavior in Fig.
\ref{Quevz}. It is evident that although for some arbitrary values
of free parameters, Quevedo's approach leads to consistent result (see Fig. %
\ref{Quevz}), one may adjust free parameters to find at least one extra
divergency. It is clear that such an extra divergency comes from the
possible real root of the following relation which is observed in
denominator of Quevedo's Ricci scalar
\begin{equation}
S \frac{dM(S,Q)}{dS}+Q \frac{dM(S,Q)}{dQ}=0.
\end{equation}

In addition, regarding the up panels of Fig. \ref{Quevz}, we find that there
is no consistency for the root of $C_{Q,P}$. So, one can regard another
consistent thermodynamical metric.

%%%%%%%%%%%%%%%%%%%%%%%%%%%%%%%%%%%%%%%%%%%%%%%%%%%%%
\begin{figure}[tbp]
$%
\begin{array}{ccc}
\epsfxsize=5.5cm \epsffile{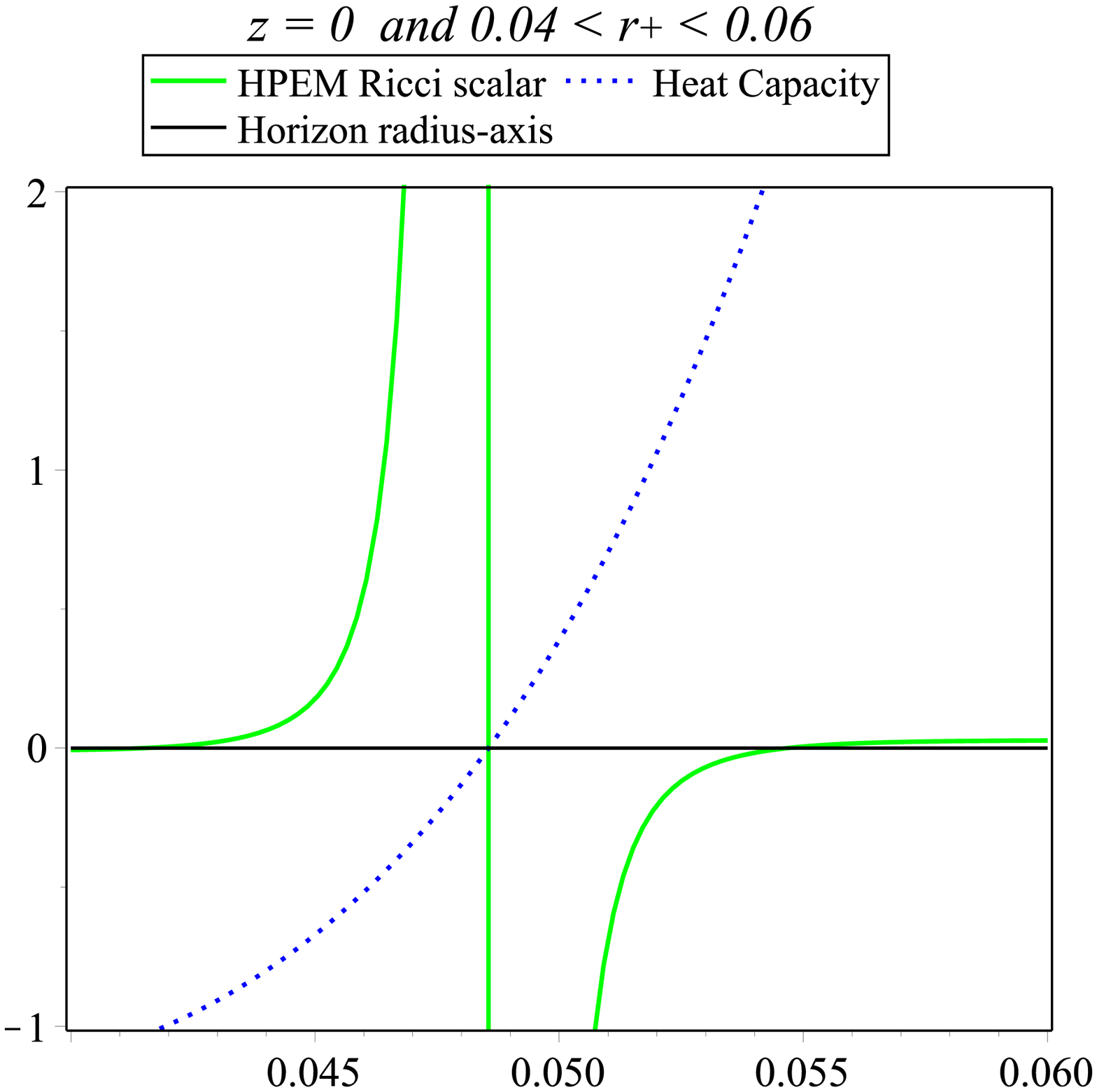} & \epsfxsize=5.5cm %
\epsffile{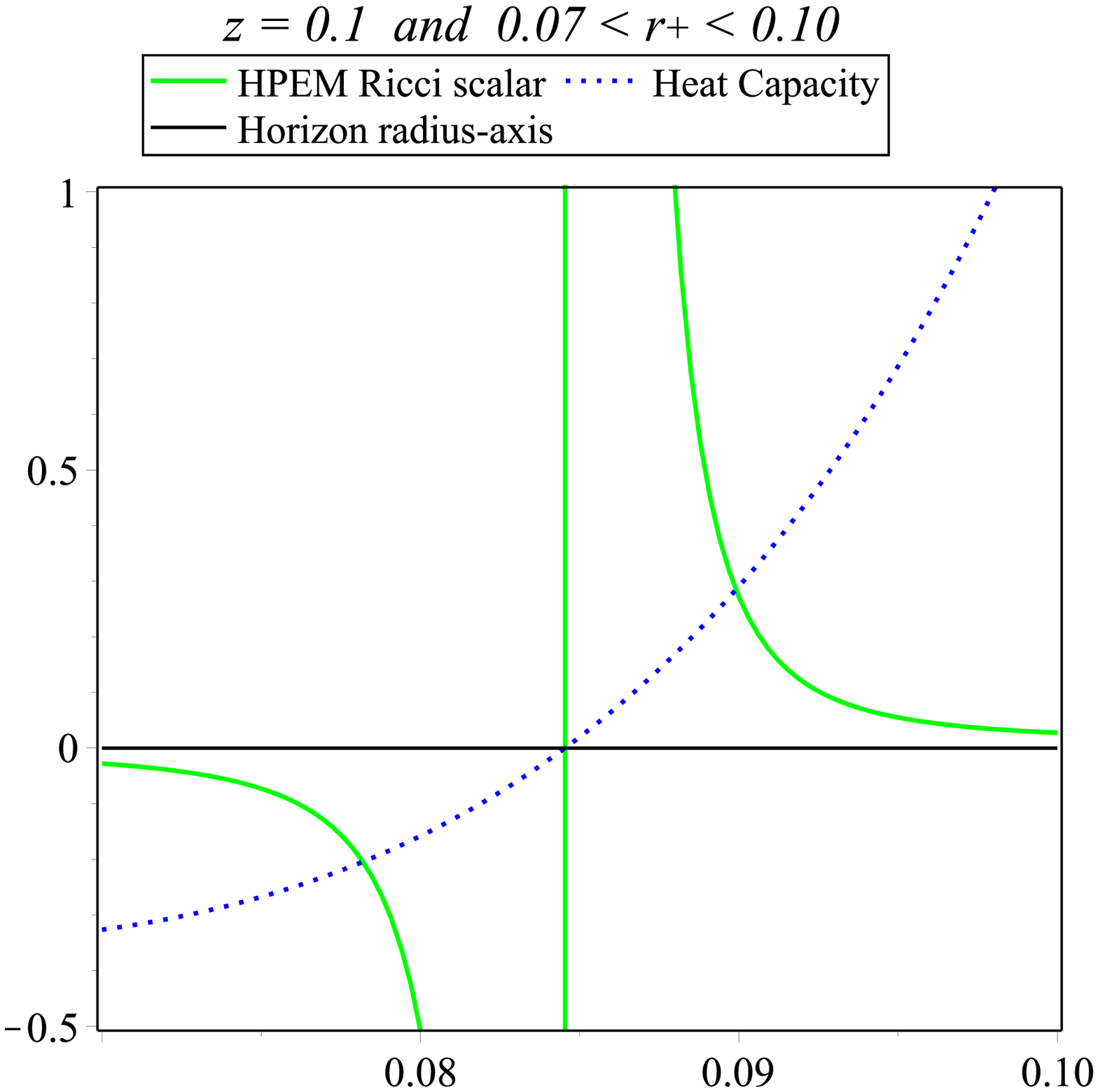} & \epsfxsize=5.5cm \epsffile{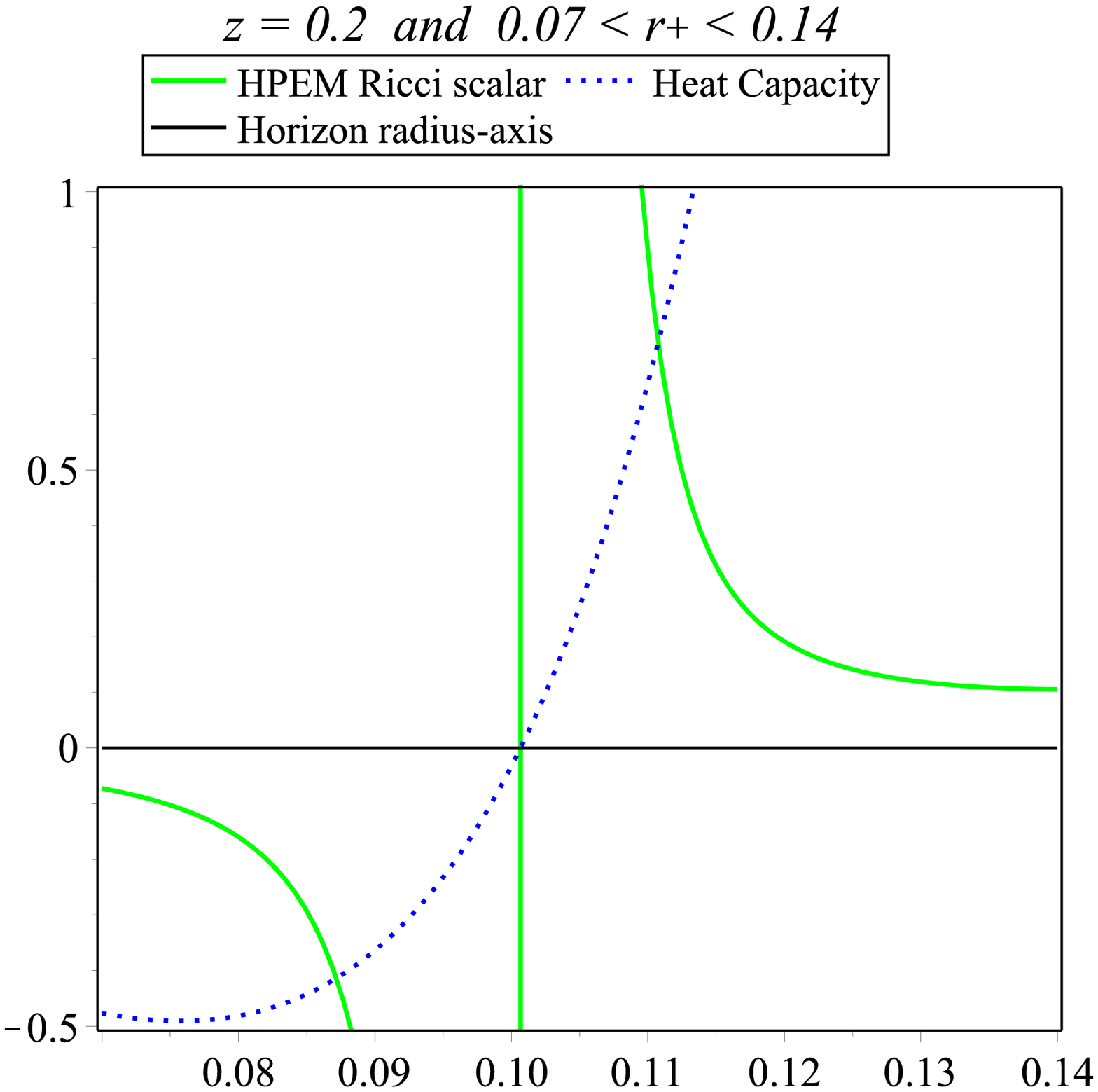} \\
\epsfxsize=5.5cm \epsffile{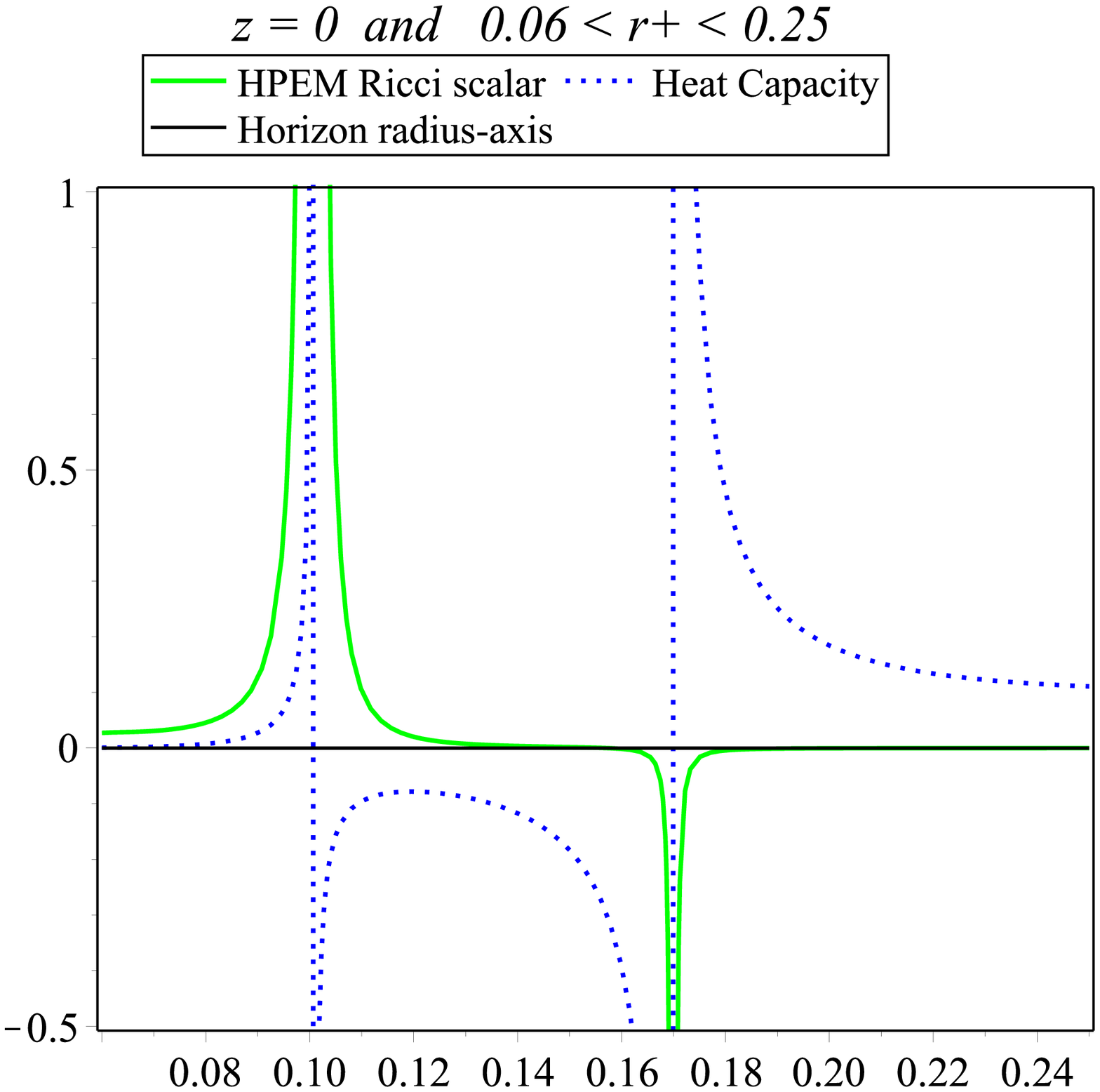} & \epsfxsize=5.5cm %
\epsffile{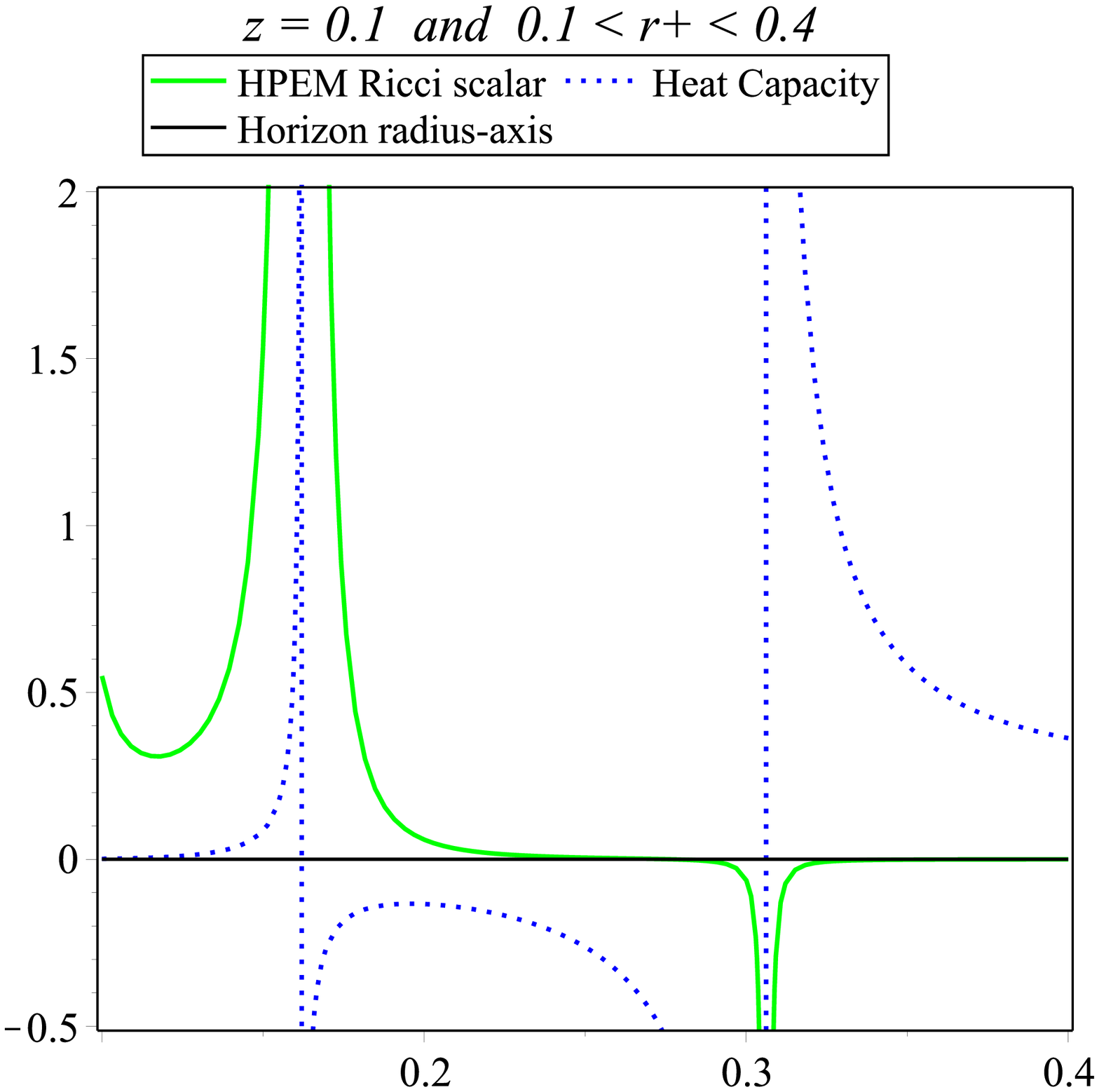} & \epsfxsize=5.5cm \epsffile{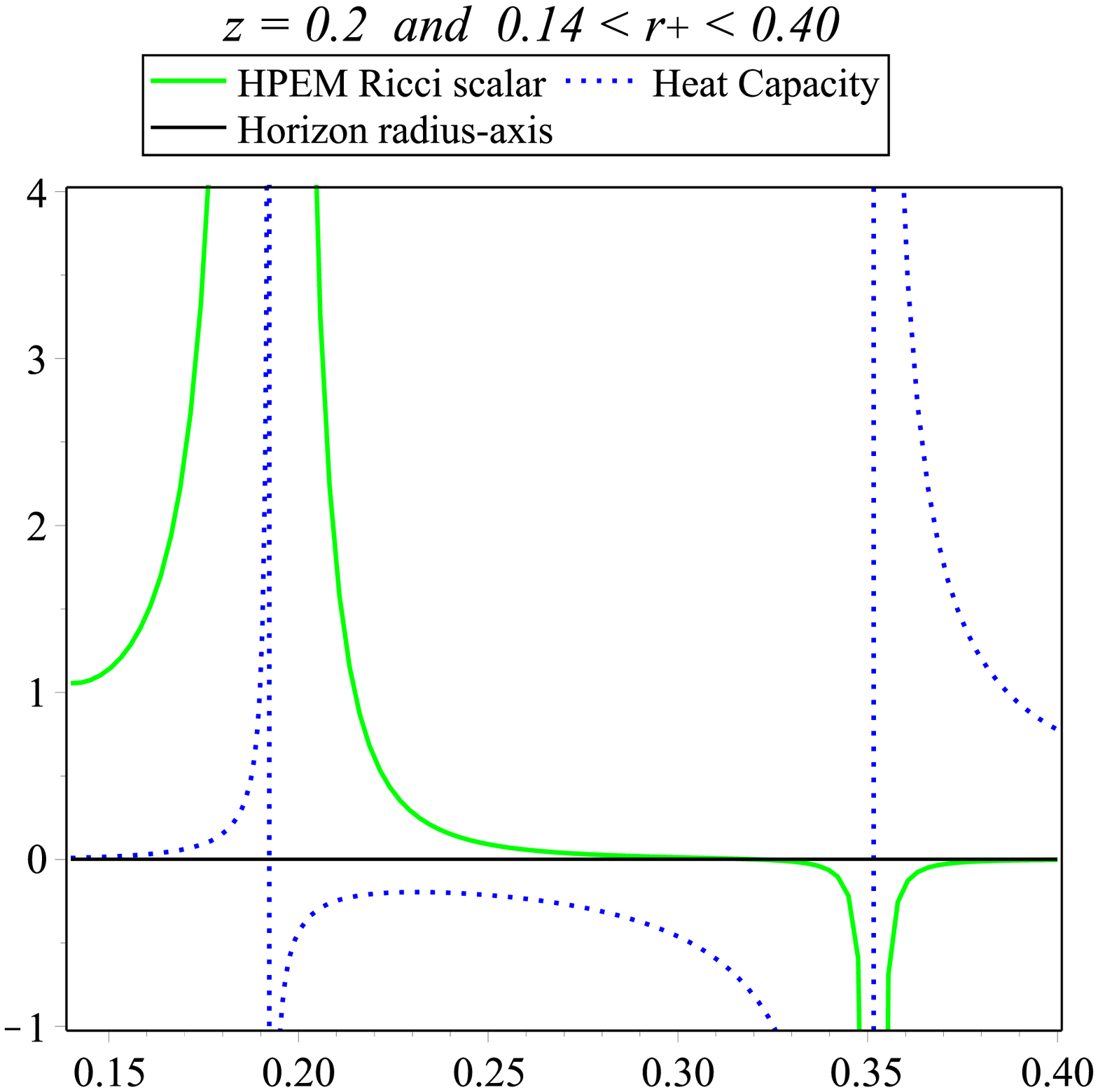}%
\end{array}
$%
\caption{Heat Capacity (dotted line) and the Ricci scalar of HPEM metric
(solid line) versus $r_{+}$ for $k=1$, $\protect\lambda=1$, $\protect\gamma%
=1 $, $q=0.05$, $V_{0}=1$, $r_{0}=1$ and $P<P_{c}$. \textbf{Left panels:} $%
z=0$, \textbf{Middle panels:} $z=0.1$, \textbf{Right panels:}
$z=0.2$ \emph{\textbf{(For more clarifications we used different
scales).}}} \label{HPEMz}
\end{figure}
%%%%%%%%%%%%%%%%%%%%%%%%%%%%%%%%%%%%%%%%%%%%%%%%%%%

Finally, we employ the HPEM metric which has a different structure from
other metrics as the following form
\begin{equation}
ds_{HPEM}=\frac{SM_{S}}{M_{QQ}^{3}}(-M_{SS}dS^{2}+M_{QQ}dQ^{2}).
\label{HPEM}
\end{equation}

It can be seen that the HPEM metric is the same as Quevedo's one
up to a conformal factor, and this fact guarantees the Legendre
invariancy of HPEM metric. Calculations show that the denominator
of the Ricci scalar of this metric is simplified as
\begin{equation}
denom(R_{HPEM})=2S^{3}M_{SS}^{2}M_{S}^{3}.  \label{DHPEM}
\end{equation}

According to Eq. (\ref{DHPEM}), one finds two sets of divergencies. One of
them ($M_{S}=0$) related to the root of temperature and the other ($M_{SS}=0$%
) corresponds to divergence points of the heat capacity. For more
clarifications, we plot Fig. \ref{HPEMz}. In this figure, one finds
different behaviors of the HPEM Ricci scalar before and after the divergence
points and root (zero $T$) of the heat capacity. In other words, the
divergence points of the Ricci scalar related to the root of heat capacity
could be distinguished from the divergencies related to phase transition
points based on the curvature scalar behavior.

\section{Conclusion}

The scalar-tensor theory of gravity is one of the well-known alternative
theories of Einstein gravity which is motivated by the low-energy limit of
superstring theory. This theory can be represented in the two conformally
related (equivalent) frames, the Jordan frame in which the scalar field is
non-minimally coupled to the metric tensor and the Einstein frame in which
the scalar field is minimally coupled to the metric tensor.

In this paper, we considered one of the special models of this theory in the
Einstein frame and investigated its Lifshitz like black hole solutions with
different horizon topologies. We found that these solutions have an
asymptotically Lifshitz like behavior.

Then, we studied thermodynamic properties of the solutions in the
extended phase space and found that these black holes may behave
like a van der Waals system depending on the choice of the
Lifshitz parameter. In other words, one can find a first order
phase transition for $T<T_c$ in isotherm $P-V$ diagram (or
correspondingly for $P<P_c$ in $G-T$ plots). We also numerically
calculated the critical quantities and showed that these critical
quantities depend on the Lifshitz parameter. Interestingly, we
found a van der Waals like phase transition for topological black
holes with planar and hyperbolic event horizons.

In order to investigate thermal stability, we calculated the heat capacity
and found that the results are completely in agreement with our previous
conclusions in the extended phase space. For the sake of completeness, we
investigated the phase transition and critical points by use of different
models of geometrical thermodynamics and interpreted their results.

It is interesting to investigate the conformally related black hole
solutions in the Jordan frame and compare their thermodynamic properties
with our results. We should note that although both Einstein and Jordan
frames are equivalent, the conformally related black hole solutions may have
different properties. In addition, one can apply metric perturbations to
investigate dynamical stability of the solutions. Moreover, one may follow
the work of Ref. \cite{Biro} and investigate new aspect of thermodynamic
properties of these black holes by regarding the Hawking radiation to define
an alternative Smarr relation with new pressure, volume and entropy. We
leave these subjects for the future work.

%%%%%%%%%%%%%%%%%%%%%%%%%%%%%%%%%%%%%%%%%%%%%%%%%%%%%%%%%%%%%%%%%%%%%%%%%%%%%%%%%%%%%%%%%%%%%%%%%%

\section*{Acknowledgements}

We would like to thank the anonymous referees for their valuable
comments which improved the quality of this manuscript
significantly. We wish to thank Shiraz University Research
Council. SHH would like to thank the hospitality of the Institute
of Theoretical Physics, Faculty of Mathematics and Physics,
Charles University during his short visit.

%%%%%%%%%%%%%%%%%%%%%%%%%%%%%%%%%%%%%%%%%%%%%%%%%%%%%%%%%%%%%%%%%%%

\bigskip

\end{document}